\documentclass[aps,prb,twocolumn,amsmath,amssymb]{revtex4}
\usepackage{graphicx}
\usepackage{amsmath}

\begin{document}

\title{Degenerate Quantum Gases with Spin-Orbit Coupling} 
\author{Hui Zhai}
\email{hzhai@tsinghua.edu.cn}
\affiliation{Institute for Advanced Study, Tsinghua University, Beijing, 100084, China}

\date{\today}

\begin{abstract}

This review focuses on recent developments on synthetic spin-orbit (SO) coupling in ultracold atomic gases. Two types of SO coupling are discussed. One is Raman process induced coupling between spin and motion along one of the spatial directions, and the other is Rashba SO coupling. We emphasize their common features in both single-particle and two-body physics and their consequences in many-body physics. For instance, single particle ground state degeneracy leads to novel features of superfluidity and richer phase diagram; increased low-energy density-of-state enhances interaction effects; the absence of Galilean invariance and spin-momentum locking give rise to intriguing behaviors of superfluid critical velocity and novel quantum dynamics; and mixing of two-body singlet and triplet states yields novel fermion pairing structure and topological superfluids. With these examples, we show that investigating SO coupling in cold atom systems can enrich our understanding of basic phenomena such as superfluidity, provide a good platform for simulating condensed matter states such as topological superfluids, and more importantly, result in novel quantum systems such as SO coupled unitary Fermi gas and high spin quantum gases. Finally we also point out major challenges and possible future directions. 

\end{abstract}
\maketitle
\tableofcontents

\section{Introduction}

The effects of SO coupling for electrons have been extensively studied in various areas of physics before. For instance, in atomic physics, it gives rise to fine structure splitting that plays a very important role in electronic structure of atoms; In condensed matter physics, it is one of the major effects that govern electron transport in a semiconductor, and recently it has been found that SO coupling can give rise to novel materials such as topological insulator, quantum anomalous Hall effect and topological superconductor.  

In contrast, neutral atoms originally do not have SO coupling effect. Nevertheless, recently by utilizing atom-light interaction, synthetic SO coupling can be generated for ultracold atoms. In the past few years, studying SO coupled quantum gas has become one of the hottest topics in cold atom physics, and many theoretical and experimental progresses have been made in this direction. One question the readers may have in mind is, since SO coupling effects have been studied in several different branches of physics before, why we want to study this effect again in the content of cold atom system and what the new physics we shall expect here. From my perspective, the main motivations for studying SO coupling effects in ultracold quantum gases of cold atoms can be summarized as follows:

\begin{itemize}

\item Previously, SO coupling effects have only been studied in fermonic systems. Whereas many atomic species are bosonic, creating SO coupling in cold atomic gases provides the first physical realization of SO coupled boson system, and therefore arises many new issues, for instance, how SO coupling affects the behavior of a boson superfluid. This also opens a new avenue where many new quantum states and novel quantum phenomena will emerge. 

\item Cold atom systems are highly tunable and can be used for the purpose of quantum simulation. Making use of this tunability, on one hand, we can study physics like topological insulator and superconductor in a more flexible and disorder free setting; and on the other hand, we can reach certain parameter regimes that are not easy to access with conventional solid state materials, for instance, tuning the strength of SO coupling to be comparable with Fermi energy, where novel effects will be expected. 

\item Cold atom systems have their own unique features, and the interplay between SO coupling and these features leads to many intriguing phenomena. For examples, using Feshbach resonance one can reach strongly interacting ``unitary" quantum gas with many universal properties. How does SO coupling affect those properties of ``unitary" gases? In contrast to spin-$1/2$ electrons, many atoms possess spin much larger than $1/2$ (for instance, Rb and Na has spin-$1$, Cr is spin-$3$, Er is spin-$6$ and Dy is spin-$8$), how does SO coupling manifest its effect in these high spin systems?

\end{itemize}

In this review, I shall illustrate above three points with examples from recent progresses, and  hopefully, it will stimulate more efforts along this direction. 

In this review, two types of SO coupling will be discussed. For the first type, spin is only coupled to motion of atoms along one spatial direction which is induced by two contour-propagating Raman beams. This has been realized in recent experiments for both bosonic and fermionic atoms \cite{Spielman_SOC,Jing_SOC,MIT_SOC}, and in this review we will call it ``Raman-induced SO coupling". The other type is Rashba SO coupling which has higher symmetry. Although it has not been realized experimentally yet, there are many theoretical proposals of how to realize it, and extensive theoretical studies have been made on this type of SO coupling. These two types of SO coupling will be discussed separately in this review because of their difference in microscopic details. Nevertheless, we will emphasize that there are quite a few common features between different types of SO coupling, which yield similar properties in many-body systems in various aspects.

So far there are already several reviews on this subject. The earliest review paper by Dalibard et al. focuses on the general idea and various schemes of how to create synthetic gauge field in cold atom system \cite{review_1}, and the non-abelian gauge field generates the effect of SO coupling. Then, a review paper by myself introduces some early works on many-body physics of ultracold atom gases with SO coupling \cite{review_2}. Recently, Galitski and Speilman published a nice review on Nature which focuses on experimental realization of SO coupling in cold atom systems and its connection to condensed matter physics \cite{review_3}. Goldman et al. give a comprehensive review of various realizations of synthetic gauge field so far, as well as interesting many-body physics for both bosonic and fermionic gases with gauge fields \cite{review_4}. In this review we shall try to minimize the overlap with the content that have been discussed in above review articles, and will refer readers to the corresponding parts of these articles for the overlap part. This review will also be restricted to SO coupling effect, and will not discuss some other developments in a more general framework of synthetic gauge fields with cold atoms, including trying to realizing large synthetic magnetic field or  dynamical gauge fields. 

\section{Realization of Spin-Orbit Coupling}

In this section, we shall first introduce two types of SO coupling that have been discussed mostly in current literatures. 

\subsection{Raman-induced SO Coupling \label{Raman_SO}}

The idea of generating SO coupling via Raman coupling has been discussed by serval earlier works \cite{Stamper-Kurn,Ian,Xiongjun}, and is first experimentally realized in Ref. \cite{Spielman_SOC}. We first consider alkali atoms like ${}^{87}$Rb and ${}^{40}$K, whose ground state electronic structure is ${}^2S_{1/2}$. The spin of these atoms ${\bf F}$ is the sum of electron spin ${\bf S}$ and nuclear spin ${\bf I}$. For instance, for ${}^{87}$Rb, $S=1/2$ and $I=3/2$. The total spin states therefore are separated by hyperfine coupling into two manifolds with $F=1$ and $F=2$, respectively \cite{Pethick}. All spin states are labelled by $|F, F_m\rangle$. In this type of experiments, we usually take a mixture of two spin states $|F,F_m\rangle$ and $|F,F_m-1\rangle$, and regard it as a pseudo-spin-$1/2$ system. For ${}^{87}$Rb, we usually take states $|1,0\rangle$ and $|1,-1\rangle$ \cite{Spielman_SOC}.

\begin{figure}[t]
\includegraphics[width=3.4 in]
{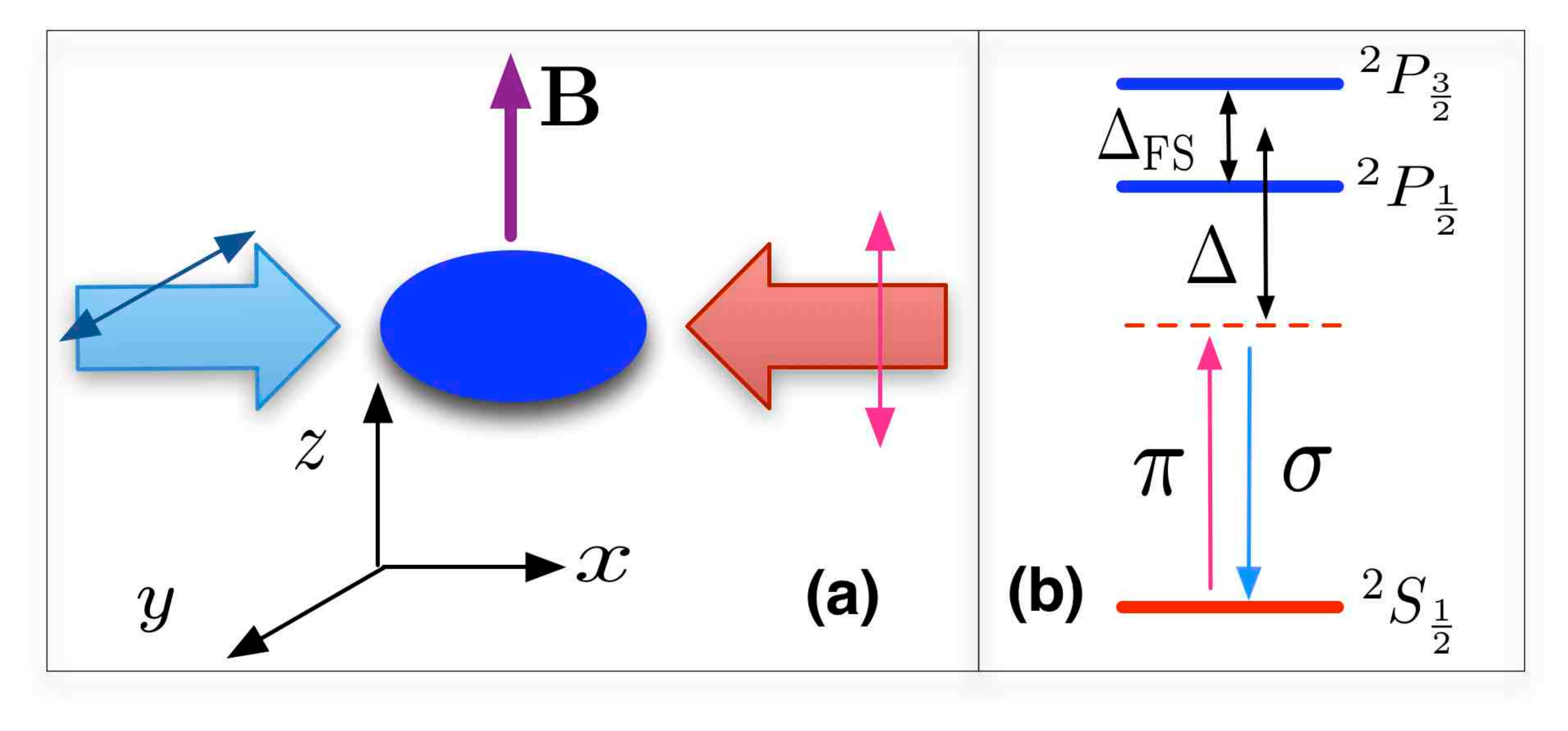}
\caption{(a) Schematic of Raman coupling scheme. (b) Atomic energy level diagram for two-photon Raman transition in alkali atom. \label{Raman}}
\end{figure}

As shown in Fig. \ref{Raman} (a), this type of SO coupling is induced by two counter-propagating Raman laser lights along $\hat{x}$. A magnetic filed along $\hat{z}$ sets the spin quantization axes. One of the laser beam is $\pi$ polarized along $\hat{z}$ and the other is linearly polarized along $\hat{y}$, and the later can be decomposed as $\sigma^+$ and $\sigma^-$. In the presence of these two laser beams, atom will undergo a two-photon process, i.e. first be excited to an intermediate excited state ${}^2P_{1/2}$ or ${}^{2}P_{3/2}$ by absorbing a $\pi$ (or $\sigma^{+/-}$) light and then come back to the ground state spin manifold by emitting a $\sigma^{+/-}$ (or $\pi$) light, as shown in Fig. \ref{Raman}(b). This two-photon process is mathematically described by a rank-$2$ tensor which can be decomposed as a sum as irreducible scalar part, vector part and tensor part. The detail derivation of this decomposition has been nicely summarized in chapter-4 of recent review article Ref. \cite{review_4}. It has been shown that if we ignore the fine structure splitting $\Delta_{\text{FS}}$ between ${}^2P_{1/2}$ and ${}^{2}P_{3/2}$, only the scalar part is nonzero which creates a spin independent scalar potential. Intuitively, it is because the ground state spin manifold of an alkali atom has ${\bf L}=0$, while this two-photon process only changes angular momentum ${\bf L}$ by one, which can not be coupled to spin degree of freedom if without the help of ${\bf L}$-${\bf S}$ coupling. That also indicates that the vector part (which is propositional to ${\bf F}$) must be proportional to the fine structure splitting $\Delta_\text{FS}$. Therefore, the strength of spin dependent Raman coupling is proportional to $\Delta_\text{FS}/\Delta^2$, rather than $1/\Delta$ (as naively expected from a second-order perturbation theory). Alternatively speaking, this is because for the spin dependent part, the second order process through ${}^2P_{1/2}$ acquires an opposite Clebsch-Gordan coefficient as the process through ${}^2P_{3/2}$. Finally, for the similar reason, the irreducible tensor term is proportional to the hyperfine structure splitting, which is usually sufficiently small comparing to $\Delta$ and can be safely ignored for alkali atom case. 

Hence, here we mainly focus on the vector term. Denoting these two spin states as $\left|\uparrow\right\rangle$ and $\left|\downarrow\right\rangle$, the single particle motion along $\hat{x}$-direction is described by Hamiltonian $\hat{H}_0$,
\begin{equation}
\hat{H}_0=\left(\begin{array}{cc}\frac{k^2_x}{2m}+\frac{\delta}{2} & \frac{\Omega}{2}e^{2ik_0x} \\\frac{\Omega}{2}e^{-2ik_0x} & \frac{k^2_x}{2m}-\frac{\delta}{2}\end{array}\right), 
\end{equation}
where $\Omega/2$ is the strength of Raman coupling and $k_0$ is the wave vector of the laser. The off-diagonal term describes the spin flipping process accompanied by a momentum transfer of $2k_0$ along $\hat{x}$ direction, which is essentially the origin of the SO coupling effect. $\delta=\omega_z-\delta\omega$, where $\omega_z$ is the Zeeman energy difference between these two spin states and $\delta\omega$ is the frequency difference between two laser beams. To derive this Hamiltonian, we also need to assume that other spin levels are far off-resonance under the two-phonon process, for instance, by utilizing the quadratic Zeeman effect \cite{Spielman_SOC,Jing_SOC, MIT_SOC, review_2}.

To illustrate the connection with conventional terminology of SO coupling, we apply a unitary transformation to the wave function $\varphi=U\psi$, with
\begin{equation}
U=\left(\begin{array}{cc}e^{-ik_0 x} & 0 \\0 & e^{i k_0 x}\end{array}\right)
\end{equation}
and correspondingly, the Hamiltonian for $\varphi$ is changed to $U\hat{H}_0U^\dag$ and the new $\hat{H}_0$ is given by 
\begin{equation}
\hat{H}_0=\left(\begin{array}{cc}\frac{(k_x+k_0)^2}{2m}+\frac{\delta}{2} & \frac{\Omega}{2} \\\frac{\Omega}{2}& \frac{(k_x-k_0)^2}{2m}-\frac{\delta}{2}\end{array}\right). \label{H_Raman0}
\end{equation}
In literature, the momentum for wave function $\varphi$ is also called ``quasi-momentum", which is related to real momentum of wave function $\psi$ by a shift of $\mp k_0$ ($-$ for $\left|\right\uparrow\rangle$ and $+$ for $\left|\right\downarrow\rangle$) along $\hat{x}$ direction.
In terms of Pauli matrix, the Hamiltonian Eq. \ref{H_Raman0} can be written as
\begin{equation}
\hat{H}_0=\frac{(k_x+k_0\sigma_z)^2}{2m}+\frac{\delta}{2}\sigma_z+\frac{\Omega}{2}\sigma_x. \label{H_Raman}
\end{equation}
Upon a spin-rotation along $\hat{y}$ by $\pi/2$, $\sigma_x\rightarrow \sigma_z$ and $\sigma_z\rightarrow -\sigma_x$, $\hat{H}_0$ can be expressed as
\begin{equation}
\hat{H}_0=\frac{(k_x-k_0\sigma_x)^2}{2m}-\frac{\delta}{2}\sigma_x+\frac{\Omega}{2}\sigma_z,
\end{equation}
which corresponds to an equal weight mixing of Rashba ($k_x\sigma_x+k_y\sigma_y$) and Dresselhaus ($k_x\sigma_x-k_y\sigma_y$) SO coupling, in addition with a Zeeman field in the spin $xz$ plane.  

\subsection{Rashba SO Coupling \label{Rashba}}

In recent literatures of cold atom systems, many theoretical papers focus on pure Rashba SO coupling whose single-particle Hamiltonian can be written as
\begin{equation}
\hat{H}_0=\frac{(k_x-k_0\sigma_x)^2}{2m}+\frac{(k_y-k_0\sigma_y)^2}{2m}. \label{H_Rashba}
\end{equation}
So far such a SO coupling has not been realized yet. However, many theoretical proposals have been made on how to realize such a SO coupling experimentally. Basically, these proposals fall into three categories. 

\begin{itemize}

\item Dark-State Scheme: This scheme utilizes Lambda or tripod laser coupling to generate two or more degenerate dark states, and these degenerate dark states serve as pseudo-spins. Since these darks states are all dressed states, and the wave functions of these dressed state have nontrivial spatial dependence, the kinetic energy operator projected into the dark state manifold usually  possesses nontrivial abelian or non-abelian gauge field term \cite{gaugea,gaugeb,dark1,dark2,dark3,dark4,dalibard}. For certain carefully designed laser configurations, these gauge field can give rise to SO coupling of the form as Eq. \ref{H_Rashba} \cite{dalibard}. This scheme has been summarized in detail in review article Ref. \cite{review_1}. However, the problem with the dark state scheme is that there is always at least one eigenstate whose energy is lower than the dark-state manifold. Thus, when the system is initially prepared in the dark-state manifold, the collision between particle will inevitably lead to the decay into the lowest energy dressed state and the lifetime of the system of interest can be quite limited \cite{collision}.  

\item Generalized Raman Scheme: In Sec. \ref{Raman_SO} we have discussed that SO coupling along one of the spatial direction can be generated by two counter-propagating Raman beams. One can generalize this scheme to realize a Rashba-type SO coupling by using more Raman beams \cite{Rashba1,Rashba2}. Here the advantage is that with Raman scheme, SO coupling is created for the manifold of the lowest energy, and therefore there is no collisional instability. However, the challenge comes from the heating problem (as we will discuss in the end of this paper), and therefore the more laser beams, the stronger the heating rate. Beside, it has also been proposed that by fast switching the direction of two counter-propagating Raman beam, a Rashba SO coupling can also been generated from a time-averaged effect \cite{Rashba3}. 

\item Pure Magnetic Scheme: To avoid the heating problem due to Raman laser, schemes that only involve magnetic method have been proposed \cite{magnetic1,magnetic2}. In fact, back to early days of cold atom physics, it has been pointed out that when a Bose condensate is trapped in a magnetic trap, the spin of atoms are frozen to the direction of magnetic field varying in space, which gives rise to an effective gauge potential \cite{Ho_gauge}. Nevertheless, in that case, the gauge filed varies on the scale of condensate size, and therefore it is relatively weak. To create a gauge potential which can vary at the scale of inter-particle spacing, one needs a magnetic field configuration that can vary at the scale of micron. Such a situation can be achieved in an atomic chip experiment. Currently, engineering Rashba SO coupling and other topological phase using atomic chip is one of the most promising directions for the future \cite{magnetic1,magnetic2,Chip}.   

\end{itemize}

In additional to Rashba SO coupling of the form as Eq. \ref{H_Rashba}, there are also interests on systems with isotropic SO coupling in all three dimensions, whose Hamiltonian is written as
\begin{equation}
\hat{H}_0=\sum\limits_{i=x,y,z}\frac{(k_i-k_0\sigma_i)^2}{2m}. \label{H_3d}
\end{equation}
Schemes of realizing such a coupling have been proposed with both generalized Raman scheme \cite{3d} and pure magnetic scheme\cite{magnetic1}.

\section{Single-particle and Two-body Physics}

In this section I will highlight some common features between these two types of SO coupling for both single-particle and two-body physics, which will will be responsible for most of interesting many-body physics discussed in this review.

\subsection{Single-Particle Physics \label{Single}}

Both kinds of SO coupling discussed above can be written in a form as a momentum-dependent Zeeman field 
\begin{equation}
\hat{H}_0=\frac{{\bf k}^2}{2m}+\vec{h}_{{\bf k}}\cdot\vec{\sigma}. \label{Zeemank}
\end{equation}
For Raman-induced SO coupling Eq. \ref{H_Raman}, 
\begin{equation}
\vec{h}_{{\bf k}}=\{\Omega/2,0,k_xk_0/m+\delta/2\};
\end{equation}
for Rashba SO coupling Eq. \ref{H_Rashba},
\begin{equation}
\vec{h}_{{\bf k}}=\{-k_xk_0/m,-k_yk_0/m,0\};
\end{equation} 
and for three-dimensional isotropic coupling Eq. \ref{H_3d}, 
\begin{equation}
\vec{h}_{{\bf k}}=\{-k_xk_0/m,-k_yk_0/m,-k_z k_0/m\}.
\end{equation}
These single particle Hamiltonians still possess translational symmetry and thus momentum ${\bf k}$ is a good quantum number.  With SO coupling, spin is no longer a good quantum number, but for each given momentum ${\bf k}$ there is another good quantum number for single-particle Hamiltonian, namely, the helicity. Helicity ``$\pm$" denotes spin parallel or anti-parallel to Zeeman field $\vec{h}_{\bf k}$, and their energies are given by
\begin{equation}
E_{{\bf k},\pm}=\frac{{\bf k}^2}{2m}\pm |h_{{\bf k}}|.
\end{equation}
Hereafter, we shall always use the recoil energy $E_\text{r}=k^2_0/(2m)$ as energy unit. 

\begin{figure}[t]
\includegraphics[width=3.4 in]
{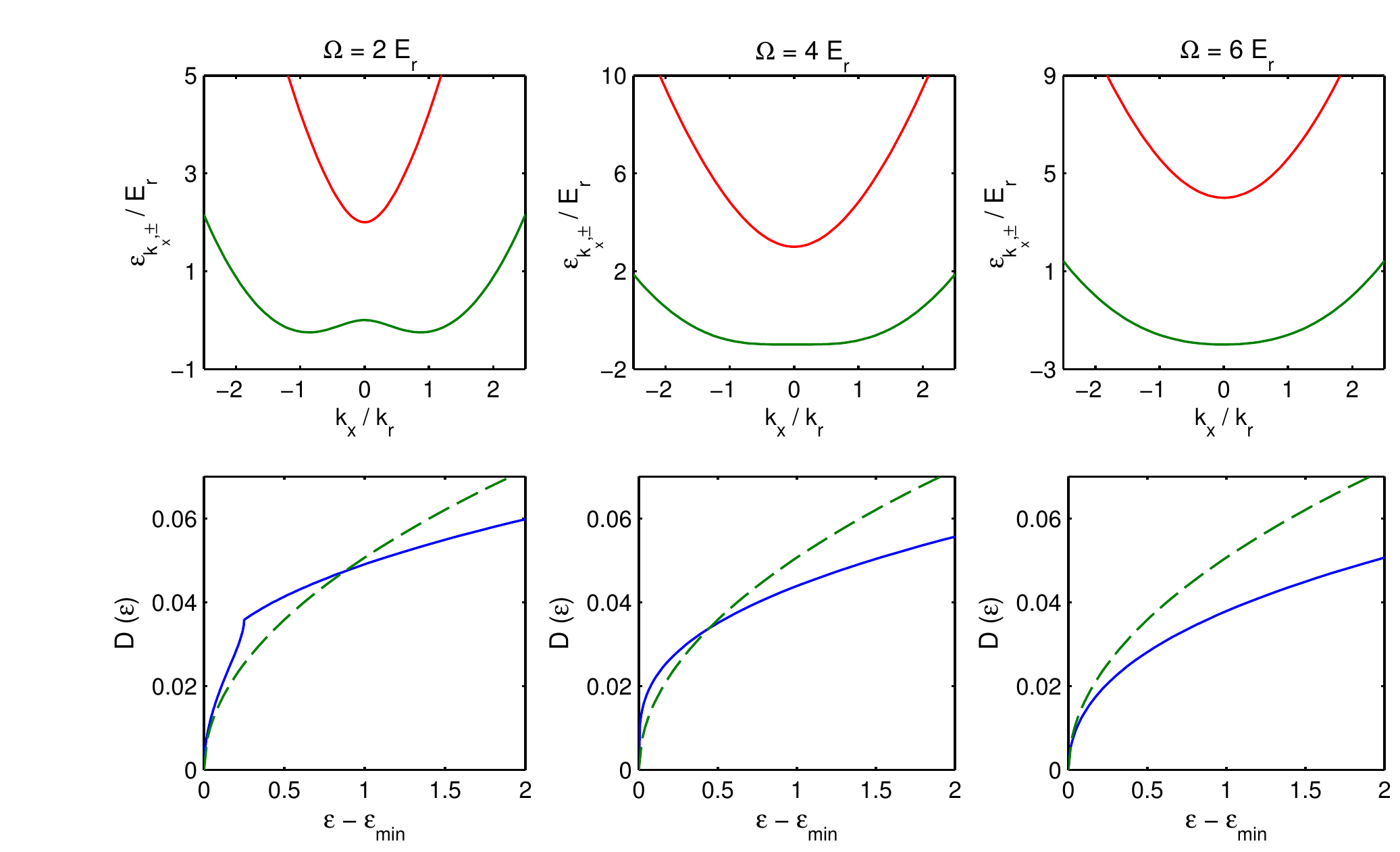}
\caption{Upper panel: single-particle dispersion for Raman-induced SO coupling with different coupling strength $\Omega$; Lower panel: corresponding single-particle density-of-state (DoS). As reference, the dashed line is DoS without SO coupling. Similar plot has been published in Ref. \cite{Zheng_Yu} \label{Single_Raman}}
\end{figure}

\vspace{0.1in}

$\circ$ \textbf {Single-particle ground state degeneracy:} For Raman-induced SO coupling, when $\delta=0$, the single-particle dispersion is given by
\begin{equation}
\epsilon_{{\bf k},\pm}=\frac{{\bf k}^2}{2m}\pm \sqrt{\frac{k^2_x k^2_0}{m^2}+\frac{\Omega^2}{4}}.
\end{equation} 
For $\Omega<4E_\text{r}$, the single-particle dispersion displays two degenerate minima at 
\begin{equation}
k_x=\pm k_0\sqrt{1-\left(\frac{\Omega}{4E_\text{r}}\right)^2}.
\end{equation}
As $\Omega$ increases toward $4E_\text{r}$, these two minima approach each other and merge into one single minimum at $\Omega=4E_\text{r}$, as shown in Fig. \ref{Single_Raman}. When $\Omega>4 E_\text{r}$, the single particle dispersion only has one single minimum at $k_x=0$.

For Rashba SO coupling, the single particle dispersion is given by
\begin{equation}
\epsilon_{{\bf k},\pm}=\frac{{\bf k}^2}{2m}\pm \frac{k_0k_\perp}{m},
\end{equation}
where $k_\perp=\sqrt{k^2_x+k^2_y}$. Thus, the single particle ground state is located at $k_\perp=k_0$ and $k_z=0$. Notice that the single particle energy is independent of the angle $\theta=\arctan(k_x/k_y)$, this Hamiltonian has a circle of degenerate ground states. 

In fact, degenerate single particle ground state is a natural consequence from SO coupling. Because the SO coupling term is usually linear in ${\bf k}$, thus, the single-particle minimum is often shifted from ${\bf k}=0$ to finite momentum. The symmetry of a finite momentum state is usually lower than that of the Hamiltonian. For instance, for the Raman-induced SO coupling with $\delta=0$, the Hamiltonian has a $Z_2$ symmetry ($k_x\rightarrow -k_x$ and $\sigma_z\rightarrow-\sigma_z$); and for Rashba SO coupling, the Hamiltonian has a $U(1)$ rotational symmetry with simultaneous rotation of spin and space in $xy$ plan along $\hat{z}$ direction. While a plane wave state with finite ${\bf k}$ does not possess these symmetries, and therefore, the ground states become degenerate. 

As we will discuss later, the degeneracy of single particle ground state causes frustration to Bose condensation, which can lead to new types of superfluid with additional broken symmetries, for instance, the condensate with spatial stripe order is first predicted by Ref. \cite{Zhai_SOCBoson} in Rashba SO coupling system, and subsequently in Raman-induced SO coupling system by Ref. \cite{Shizhong}. Large single particle degeneracy can even lead to more exotic ground state of bosons, as we shall discussed in later session. 

\vspace{0.1in}

$\circ$ \textbf{Change of density-of-state:} The change of single particle dispersion also affects low-energy density-of-state (DoS), and in many cases, the DoS is significantly enhanced. In Raman-induced SO coupling, at $\Omega=4E_\text{r}$ with $\delta=0$, the single particle dispersion exhibits a $k^4_x$ dispersion along $\hat{x}$ for small $k_x$. Thus, the DoS becomes larger than usually $k^2_x$ dispersion, as shown in Fig.  \ref{Single_Raman}. This will affect Bose condensation temperature and the interaction shift of Bose condensation temperature \cite{Zheng_Yu}. 

More dramatic change of single-particle DoS is the Rashba SO coupling and the three-dimensional isotropic SO coupling \cite{Vijay_2body,Cui_DoS,Kamenev}. For Rashba SO coupling, if the system is two-dimensional, the low-energy DoS behaves as $1/\sqrt{\epsilon}$ as normal one-dimension system; and if the system is three-dimensional, the low-energy DoS behaves as a constant as normal two-dimensional case. Effectively, this reduces the dimensionality by one. Intuitively, this dimension reduction is because the ground states form a circle in $k_xk_y$ plane, in two-(three-)dimension, energy increases only along the one-(two-)direction(s) perpendicular to the circle. For three-dimensional isotropic SO coupling, the low-energy DoS behaves as $1/\sqrt{\epsilon}$ which effectively reduces the dimensionality by two. The drastically increased DoS will has important consequences when interaction effects are included. For instance, as we will discuss in detail later, it will increase quantum fluctuation of interacting bosons, and will also enhance the tendency of pair formation in attractive Fermi gases.  

\vspace{0.1in}

$\circ$ \textbf{Absence of Galilean invariance:} As mentioned above, the system with SO coupling still displays translational invariance and momentum is still a good quantum number. However, the Galilean invariance is absent \cite{Zheng_Yu,Shuai_mode,Wu_Biao,Supercurrent}. Under the Galilean transformation, ${\bf r}\rightarrow {\bf r}-{\bf v}t$ and the Hamiltonian becomes $\hat{H}_0-{\bf v}{\bf k}$. For conventional system with $H_0={\bf k}^2/(2m)$, the additional term ${\bf v}{\bf k}$ can be eliminated by a gauge transformation $e^{i{\bf v}{\bf r}}$, and thus the Hamiltonian is invariant under the Galilean transformation. While with SO coupling, for instance, with Raman-induced type, after the same gauge transformation to eliminate the ${\bf v}\cdot{\bf k}$ term, it acquires an additional term $v_xk_0\sigma_z/m$; and for Rashba type SO coupling, it acquires an additional term $-k_0\vec{v}_\perp\cdot\vec{\sigma}_\perp/m$. That is to say, in the moving frame, the system will have an additional Zeeman term proportional  to the velocity of moving frame. Hence, both the spin wave function and the energy dispersion will change with velocity. This will manifest itself in our later discussion of collective excitation and superfluid critical velocity.  

\vspace{0.1in}

$\circ$ \textbf{Spin-momentum locking:} Since the SO coupling Hamiltonian can be written as a momentum dependent Zeeman field, it means that the spin points to different direction for eigenstates with different momentum. For instance, for the Raman-induced SO coupling, if the particle moves with large positive $k_x$, the ground state spin almost points up along $\hat{z}$; and with large negative $k_x$, spin points to opposite direction. Later we shall show that this spin-momentum locking can be directly imaged in collective modes experiments and spin-resolved spectroscopy. For Rashba SO coupling, around certain closed loop in the momentum space, the spin direction even twists a topologically nontrivial loop in the Bloch sphere that leads to topologically nontrivial states in the system.   

\subsection{Two-Body Physics \label{TB}}

 A number of theoretical works have studied two-body problem with SO coupling using different methods \cite{Cui_two-body,Peng_BP,Yu_BP,Peng_renormalization,Magarill,Vijay_2body,Yu_SR,Peng_2e, Vijay_molecule,Han_molecule,Melo_molecule,You,Blume}. In this subsection we shall discuss some common features in the two-body physics.

\vspace{0.1in}

\begin{figure}[tbp]
\includegraphics[width=0.3\textwidth]{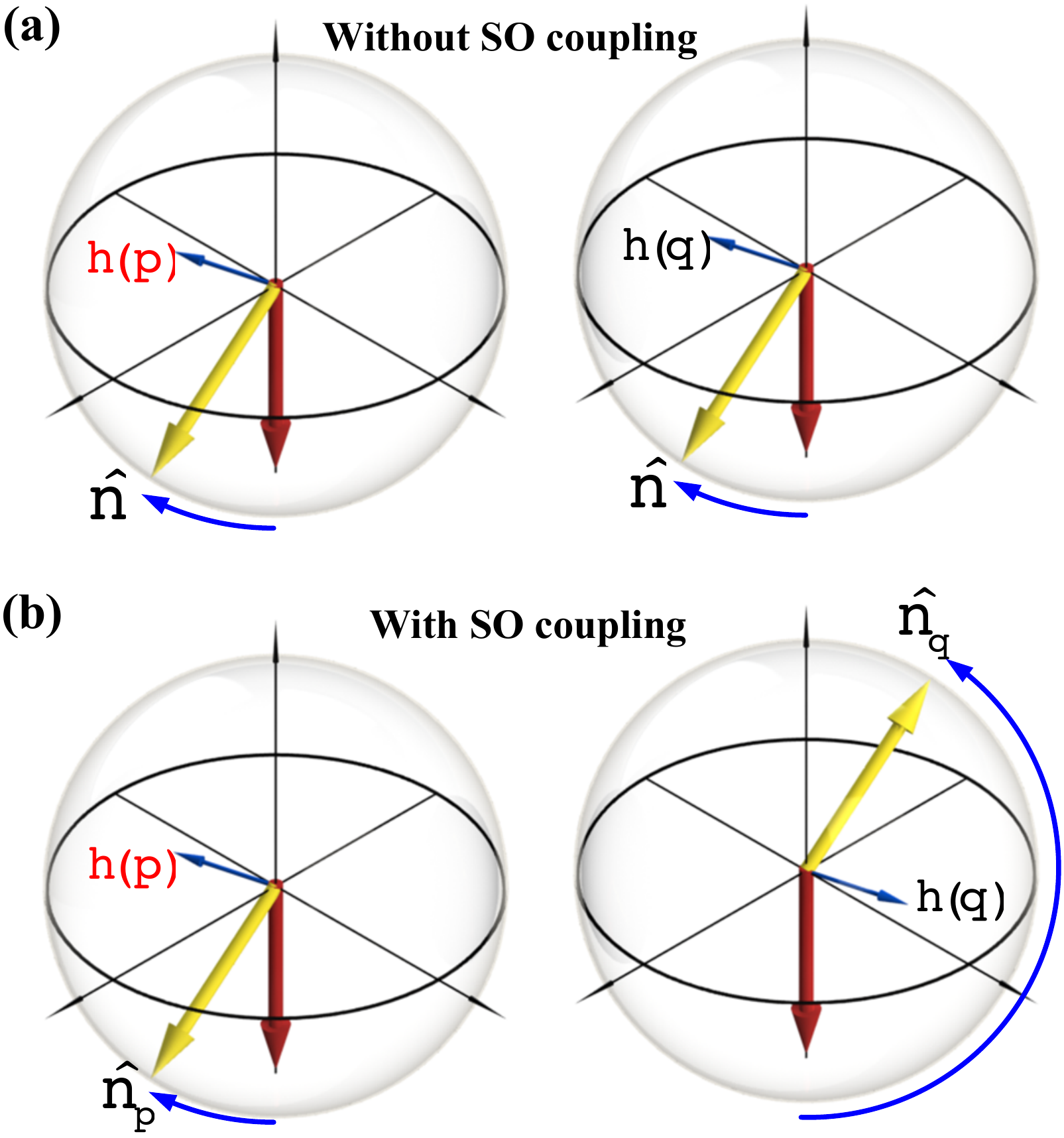}
\caption{(Color online). Schematic of how singlet and triplet states are coupled by SO coupling. Left and right column
represent spins of two atoms with different momentum. (a) represents the case in the absence of SO coupling and (b)
represents the case in the presence of SO
coupling, i.e. case (a): ${\bf h}({\bf p})={\bf h}({\bf q})$; case (b):
${\bf h}({\bf p})\neq{\bf h}({\bf q})$ when ${\bf p}\neq{\bf q}$.
Red arrows represent spin direction at $t=0$ and the yellow arrows
represent spin direction at finite time $t$. Reprinted from Ref.  \cite{Rabi} \label{ST_mix} }
\end{figure}

$\circ$ \textbf{Singlet channel and triplet channel are mixed:} Intuitively, this can be easily understood as follows \cite{Rabi}.  Let us consider two atoms initially prepared in the same spin state (say $\left|\downarrow\right\rangle$), with different momenta ${\bf p}$ and ${\bf
q}$, as represented by blue arrows in Fig. \ref{ST_mix}. The initial
state under anti-symmetrization is given by
\begin{equation}
\left|\Psi\right\rangle_{i}=\frac{1}{\sqrt{2}}\left(\left|{\bf p}\right\rangle_1\left|{\bf
q}\right\rangle_2\mp \left|{\bf q}\right\rangle_1\left|{\bf
p}\right\rangle_2\right)\left|\downarrow\right\rangle_1\left|\downarrow\right\rangle_2.
\end{equation}
where $\mp$ is for fermions and bosons, respectively. If the Zeeman-field is momentum independent (i.e. the case without SO coupling), these two atoms
with different momentum always rotate in the same way. Therefore, at
a given time $t$, two spins always rotate to the same direction
$|\hat{n}\rangle$, as shown in Fig. \ref{ST_mix}(a). The final state
wave function is then given by
\begin{equation}
|\Psi\rangle_{f}=\frac{1}{\sqrt{2}}\left(|{\bf p}\rangle_1|{\bf
q}\rangle_2\mp |{\bf q}\rangle_1|{\bf
p}\rangle_2\right)|\hat{n}\rangle_1|\hat{n}\rangle_2.
\end{equation}
Obviously this state remains as a triplet. While as shown in Eq. \ref{Zeemank}, 
SO coupling can be described as a momentum dependent Zeeman field. Thus, the amount of rotation each spin executes depends on
its momentum and is in general  different for different momentum. 
Suppose at time $t$, atom with momentum ${\bf p}$ rotates to
$|\hat{n}_{\bf p}\rangle$ and atom with momentum ${\bf q}$ rotates
to $|\hat{n}_{\bf q}\rangle$, as shown in Fig. \ref{ST_mix}b, the
final-state wave function can now be written as
\begin{equation}
|\Psi\rangle_{f}=\frac{1}{\sqrt{2}}\left(|{\bf p}\rangle_1|{\bf
q}\rangle_2|\hat{n}_{\bf p}\rangle_1|\hat{n}_{\bf q}\rangle_2\mp |{\bf
q}\rangle_1|{\bf p}\rangle_2|\hat{n}_{\bf q}\rangle_1|\hat{n}_{\bf
p}\rangle_2\right) \label{wf}.
\end{equation}
It is straightforward to show that the wave function Eq. (\ref{wf}) can
be rewritten as
\begin{align}
|\Psi\rangle_{f}=&\frac{\left(|{\bf p}\rangle_1|{\bf
q}\rangle_2\mp |{\bf q}\rangle_1|{\bf
p}\rangle_2\right)}{2}|\widetilde{T}\rangle\nonumber\\
+&\frac{\left(|{\bf
p}\rangle_1|{\bf q}\rangle_2\pm |{\bf q}\rangle_1|{\bf
p}\rangle_2\right)}{2}|\widetilde{S}\rangle \nonumber
\end{align}
where $|\widetilde{T}\rangle= (|\hat{n}_{\bf
p}\rangle_1|\hat{n}_{\bf q}\rangle_2+|\hat{n}_{\bf
q}\rangle_1|\hat{n}_{\bf p}\rangle_2)/\sqrt{2}$ and
$|\widetilde{S}\rangle= (|\hat{n}_{\bf p}\rangle_1|\hat{n}_{\bf
q}\rangle_2-|\hat{n}_{\bf q}\rangle_1|\hat{n}_{\bf
p}\rangle_2)/\sqrt{2}\propto|S\rangle$ are triplet and singlet
components, respectively. Thus, the wave function contains both singlet and triplet components. In later section we shall discuss how this mixing is probed in Raman-induced SO coupling experiment \cite{Rabi}.

\vspace{0.1in}

$\circ$ \textbf{Different partial wave channels are mixed:} This is a natural consequence that SO coupling breaks spatial rotational symmetry. Experimentally, by colliding two SO coupled Bose condensates, the scattering halo reflects the behavior of scattered wave function at large distance. For two-component bosons, contributions from $d$- and $g$-wave have been observed in the scattering halo \cite{Spielman_partial}. However, this should be distinguished from short-range behavior of two-body wave function. In usual scattering theory, wave function in different partial wave channel has different behavior at short distance. One natural question is whether and how the SO coupling changes the short-range behavior of the two-body wave function.   

For $s$-wave interaction, it is well known that the lowest energy two-body wave function takes the form as $1/r-1/a_\text{s}$ at short distance, where $a_\text{s}$ is the $s$-wave scattering length. It has been shown that when the width of square well $r_0$ is negligible comparing to $1/k_0$ ($k_0$ denotes the typical strength of SO coupling), and the scattering volume of $p$-wave is also negligible (i.e. away from $p$-wave resonance), the short-range behavior for $s$-wave at singlet channel remains as $1/r-1/a_\text{s}$, and the correction due to SO coupling is at higher partial wave channel ($p$-wave for SO couplings under current consideration) and does not contain any singular contribution term \cite{Cui_two-body,Peng_BP,Yu_BP}. This means that we can apply the same momentum space renormalization scheme to treat $s$-wave zero-range interaction even in the presence of SO coupling  \cite{Peng_renormalization}. 

\vspace{0.1in}

$\circ$ \textbf{Density-of-state effect changes low-energy scattering and the bound state formation:} The DoS effect is model dependent. For instance, with Rashba SO coupling, the tendency of forming a bound state is dramatically enhanced \cite{Magarill, Vijay_2body}. In three-dimension, there is always a shallow zero-momentum bound state for any scattering length because the DoS becomes the same as a conventional two-dimensional system \cite{Vijay_2body}. The two-body short-range correlation is also enhanced \cite{Yu_SR}. In two-dimension, for zero-momentum scattering the scattering amplitude algebraically depends on energy, similar as conventional one-dimensional system \cite{Peng_2e}, which is also due to the increased joint DoS. Similar effect can be found for an isotropic SO coupling in three-dimension, in which the DoS is increased to that of a conventional one-dimensional system \cite{3d,Vijay_2body}. 

While for Raman-induced SO coupling, because the Raman coupling term effectively plays a role as a Zeeman-field along $\hat{x}$-direction, the two-body joint DoS is in fact suppressed, and the tendency of forming bound state is suppressed. The threshold for the appearance of bound state is shifted to the positive $a_\text{s}$ side \cite{Peng_renormalization,Vijay_molecule,Han_molecule,Melo_molecule}. This has been verified experimentally and will be discussed in detail in later section. 

\section{Raman-induced Spin-Orbit Coupling}

In this section we will discuss interacting bosons and fermions with Raman-induced SO coupling, whose single-particle Hamiltonian is given by Eq. \ref{H_Raman}. So far several experimental groups have studied cold atoms in this setup with $^{87}$Rb, $^{40}$K and $^{6}$Li. We shall emphasize these experimental progresses in this direction. 

\subsection{Bosons}

\subsubsection{Phase Diagram}

In this part we will first discuss equilibrium ground state for bosons with Raman-induced SO coupling. The emphasis is that the single particle degeneracy will lead to nontrivial superfluid phase, which breaks extra symmetry in addition to the normal $U(1)$ phase symmetry.

As discussed in previous section \ref{Single}, for Hamiltonian Eq. \ref{H_Raman}, there are two degenerate minima for $\delta=0$. For nonzero $\delta$, even these two minima are not exactly degenerate, there are still two local minima in single-particle spectrum for a wide range of small $\delta$. The locations of these minima are denoted by ${\bf k}_{\pm}$ and their wave functions are given by

\begin{equation}
\varphi (\mathbf{k})=e^{i\mathbf{k}\cdot \mathbf{r}}\left( 
\begin{array}{c}
\cos \theta _{\mathbf{k}} \\ 
-\sin \theta _{\mathbf{k}}%
\end{array}%
\right)
\end{equation}%
with 
\[
\sin ^{2}\theta _{\mathbf{k}}=\frac{1}{2}%
\left( 1-\frac{k_{x}k_{0}/m+\delta /2}{\sqrt{\left( k_{x}k_{0}/m+\delta
/2\right) ^{2}+\Omega ^{2}/4}}\right), 
\]%
and their energy is given by
\begin{equation}
\epsilon_{{\bf k}}=\frac{{\bf k}^2}{2m}-\sqrt{\left(\frac{k_xk_0}{m}+\frac{\delta}{2}\right)^2+\frac{\Omega^2}{4}}
\end{equation}
Without loss of generality, we can assume the condensate wave function take the form as  \cite{Spielman_SOC,Shizhong,Stringari_phasediagram, Zheng_Yu}
\begin{equation}
\psi =\cos \alpha e^{i\theta/2} \varphi(\mathbf{p_{+}})+\sin \alpha e^{-i\theta/2} \varphi(\mathbf{%
p_{-}}),  \label{vari}
\end{equation}
where the superposition coefficient $\alpha$ and $\theta$ are taken as variational parameters. 
If the interaction is $SU(2)$ spin-rotational invariant, the interaction energy is independent of momentum, ${\bf p}_{\pm}$ can be fixed at single-particle minimum ${\bf k}_{\pm}$. However, in $^{87}$Rb system studied in current experiment, it is a pseudo-spin-$1/2$ system for bosons and the interaction term does not require any $SU(2)$ symmetry. Generally the interaction term is given by
\begin{equation}
\epsilon_\text{int}=\frac{1}{2}\int d^3{\bf r}\left(g_{\uparrow\uparrow}n^2_{\uparrow}+2g_{\uparrow\downarrow}n_{\uparrow}n_{\downarrow}+g_{\downarrow\downarrow}n^2_\downarrow\right).
\end{equation} Moreover, because of the spin-momentum locking discussed in Sec. \ref{Single}, the spin wave function is momentum dependent and thus the single-particle self-energy also becomes momentum dependent. Hence, in the ansatz Eq. \ref{vari}, ${\bf p}_{\pm}$ is taken as another variational parameters \cite{Stringari_phasediagram}. All these variational parameters should be fixed by interactions. 

\begin{figure}[t]
\includegraphics[width=3.0 in]
{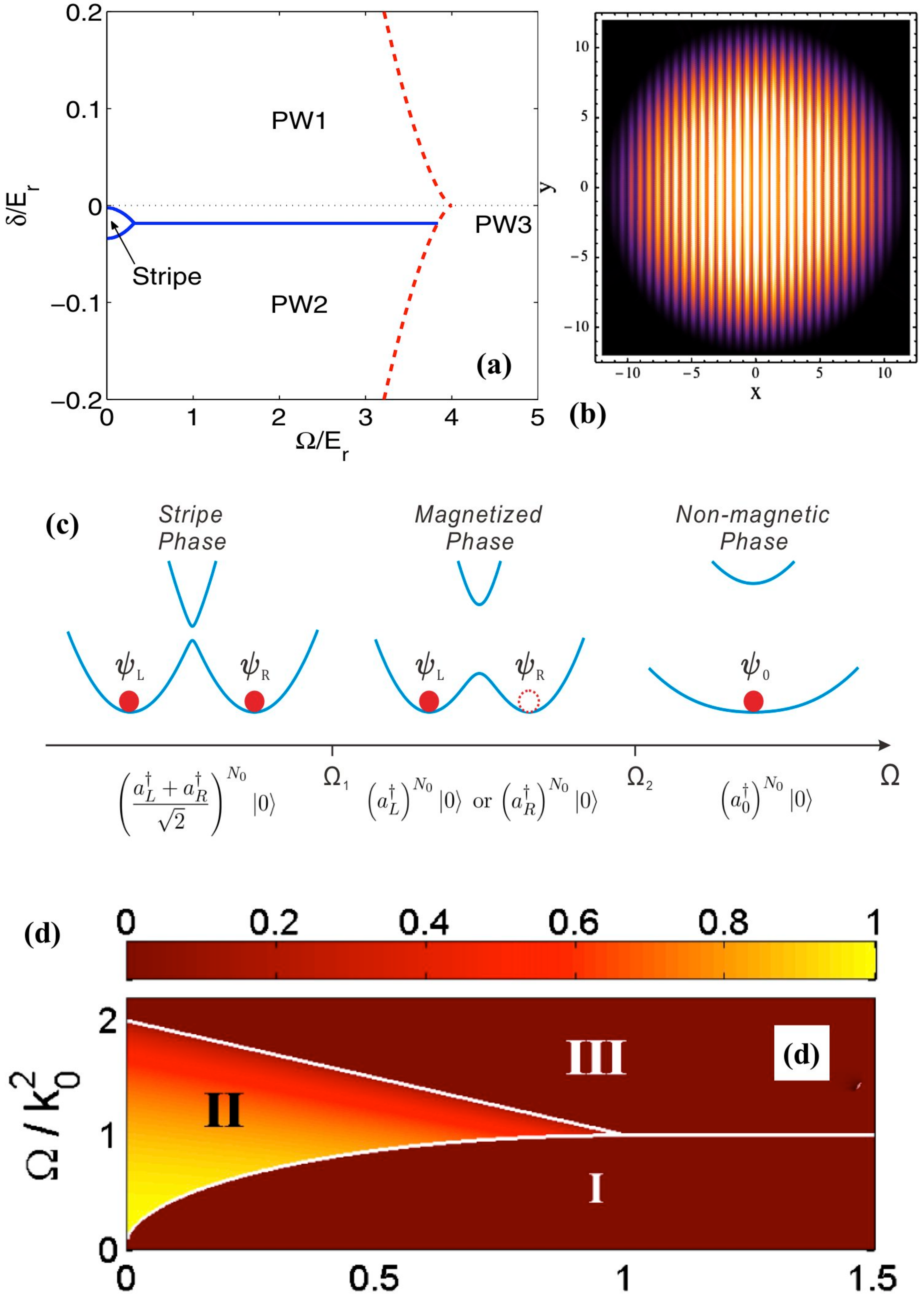}
\caption{(a) Phase diagram in terms of $\delta$ and $\Omega$ for $^{87}$Rb, reprinted from Ref. \cite{Zheng_Yu}. (b) Schematic of density distribution of stripe phase, reprinted from Ref. \cite{Shizhong}. (c) Phase diagram in term of $\Omega$ with $\tilde{\delta}=0$, condensate wave function and symmetry breaking properties are also shown schematically, reprinted from Ref. \cite{Shuai_FiniteT}. (d) Phase diagram in terms of both density $n$ and $\Omega$ with $\tilde{\delta}=0$ reprinted from Ref. \cite{Stringari_phasediagram}, where $I$, $II$ and $III$ correspond to ``Stripe Phase", ``Magnetized Phase" and ``Non-magnetic Phase" in (c), respectively, \label{Raman_Phase_Diagram}}
\end{figure}

For $^{87}$Rb atom used in current experiment, we denote $|F=1,m_\text{F}=0\rangle$ state as $\left|\uparrow\right\rangle$ and $|F=1,m_\text{F}=-1\rangle$ as $\left|\downarrow\right\rangle$, the interaction parameter $g_{\uparrow\uparrow}=c_0$, $g_{\uparrow\downarrow}=g_{\downarrow\downarrow}=c_0+c_2$, where $c_0=7.79 \times 10^{-12} \text{Hz}\text{cm}^3$, and $c_2=-3.61\times 10^{-14}\text{Hz}\text{cm}^3$. With these interaction parameters and typical density of a Bose condensate, the energy minimization gives rise to a phase diagram shown in Fig. \ref{Raman_Phase_Diagram}(a). If the minimization yields $0<\alpha<\pi/2$, the condensate wave function contains two components with different momenta (${\bf p}_{+}$ and ${\bf p}_{-}$), and therefore, spatially it will display a periodically density modulation due to the interference of these two components, as shown in Fig. \ref{Raman_Phase_Diagram}(b). This phase is denoted as ``Stripe Phase" in the phase diagram Fig. \ref{Raman_Phase_Diagram}(a). The density modulation in the stripe phase increases as $\Omega$ increases, and thus, the density-density interaction energy cost increases. Thus, above certain $\Omega$, the energy minimization will yield $\alpha=0$ or $\alpha=\pi/2$, the condensate wave function becomes a plane wave, either with momentum ${\bf p}_{+}$ or with momentum ${\bf p}_{-}$, which is denoted as ``Plane Wave Phase" (``PW1" and ``PW2" in the phase diagram Fig. \ref{Raman_Phase_Diagram}(a)). Furthermore, as Raman coupling strength $\Omega$ further increases, ${\bf p}_{+}$ and ${\bf p}_{-}$ approach each other and finally they merge together. This regime is denoted by ``PW3" in the phase diagram Fig. \ref{Raman_Phase_Diagram}(a).  

To discuss symmetry properties of these phases, we shall first consider following simplification of the Hamiltonian. Denoting $g_{\uparrow\uparrow}=g+\delta g$ and $g_{\uparrow\uparrow}=g-\delta g$, there will be a term in the interaction energy proportional to $\delta g(n^2_\uparrow-n^2_\downarrow)=\delta g (n_\uparrow-n_\downarrow)(n_\uparrow+n_\downarrow)$. Approximately, by replacing $n_\uparrow+n_\downarrow$ with $\bar{n}$, this term can be absorbed in the detuning $\delta$-term by redefining $\tilde{\delta}=\delta+\delta g\bar{n}$. Then, the remaining interaction terms are also simplified as
\begin{equation}
\epsilon_\text{int}=\frac{1}{2}\int d^3{\bf r}\left(gn^2_\uparrow+2g_{\uparrow\downarrow}n_\uparrow n_\downarrow+gn^2_\downarrow\right)
\end{equation}
Now considering the situation with $\tilde{\delta}=0$, the Hamiltonian possesses a $Z_2$ symmetry, that is, ${\bf k}\rightarrow -{\bf k}$ and simultaneously $\sigma_z\rightarrow -\sigma_z$.

In this situation, the many-body wave function for a stripe phase is written as 
\begin{equation}
\Psi=\left(\frac{e^{i\theta/2}\psi^\dag_{\text{L}}+e^{-i\theta/2}\psi^\dag_\text{R}}{\sqrt{2}}\right)^{N_0}|0\rangle. \label{WF_stripe}
\end{equation} 
where $\psi^\dag_\text{R}$ and $\psi^\dag_\text{L}$ are creation operators for single particle state with momentum ${\bf p}_{+}$ and ${\bf p}_{-}$, respectively. In addition to the normal $U(1)$ global phase symmetry breaking, this state breaks spatial translational symmetry, that is, the relative phase $\theta$ can spontaneously choose any value between zero and $2\pi$. When $\Omega>\Omega_1$, the system enters the plane wave phase, in which the many-body wave function is either $\psi^{\dag N_0}_\text{L}|0\rangle$ or $\psi^{\dag N_0}_\text{R}|0\rangle$. In this case, the ground state does not break spatial translational symmetry but breaks the $Z_2$ symmetry. Since the single particle states with opposite momentum display opposite magnetization, the ground state will also possess finite magnetization along $\hat{z}$ once the $Z_2$ symmetry is broken. Thus, it is also called ``Magnetized Phase" in the phase diagram shown in Fig. \ref{Raman_Phase_Diagram}(c) for $\tilde{\delta}=0$ case. Finally, when $\Omega>\Omega_2$, two energy minima merge into one single minimum at zero-momentum. Thus, bosons all condense into zero-momentum state as normal Bose condensate, which does not break any additional symmetry and possesses zero magnetization along $\hat{z}$. Thus, as one can see, the ground state degeneracy gives rise to novel superfluids with extra symmetry breaking properties and rich phase diagram at zero-temperature, as summarized in Fig. \ref{Raman_Phase_Diagram}(c). The mean-field phases have also been generalized to spin-1 case \cite{Lan} and double layer BEC with SO coupling in different directions \cite{Sun}, where more richer structures have been found.  

In $^{87}$Rb case since $c_2$ is quite small comparing to $c_0$ and $g_{\uparrow\downarrow}$ are very close to $g$, for typical density $\Omega_1$ is about $0.2 E_\text{r}$ which is quite small. Imaging one can increase density to increase the effect of interactions, as shown in Ref. \cite{Stringari_phasediagram}, $\Omega_1$ increases with the increasing of density, and $\Omega_2$ decreases with the increasing of density. Finally at certain critical density they meet at a tricritical point, above which they will be a direct transition from stripe phase to zero-momentum non-magnetic phase. The phase diagram in term of density is shown in Fig. \ref{Raman_Phase_Diagram}(d).

\begin{figure}[t]
\includegraphics[width=3.2 in]
{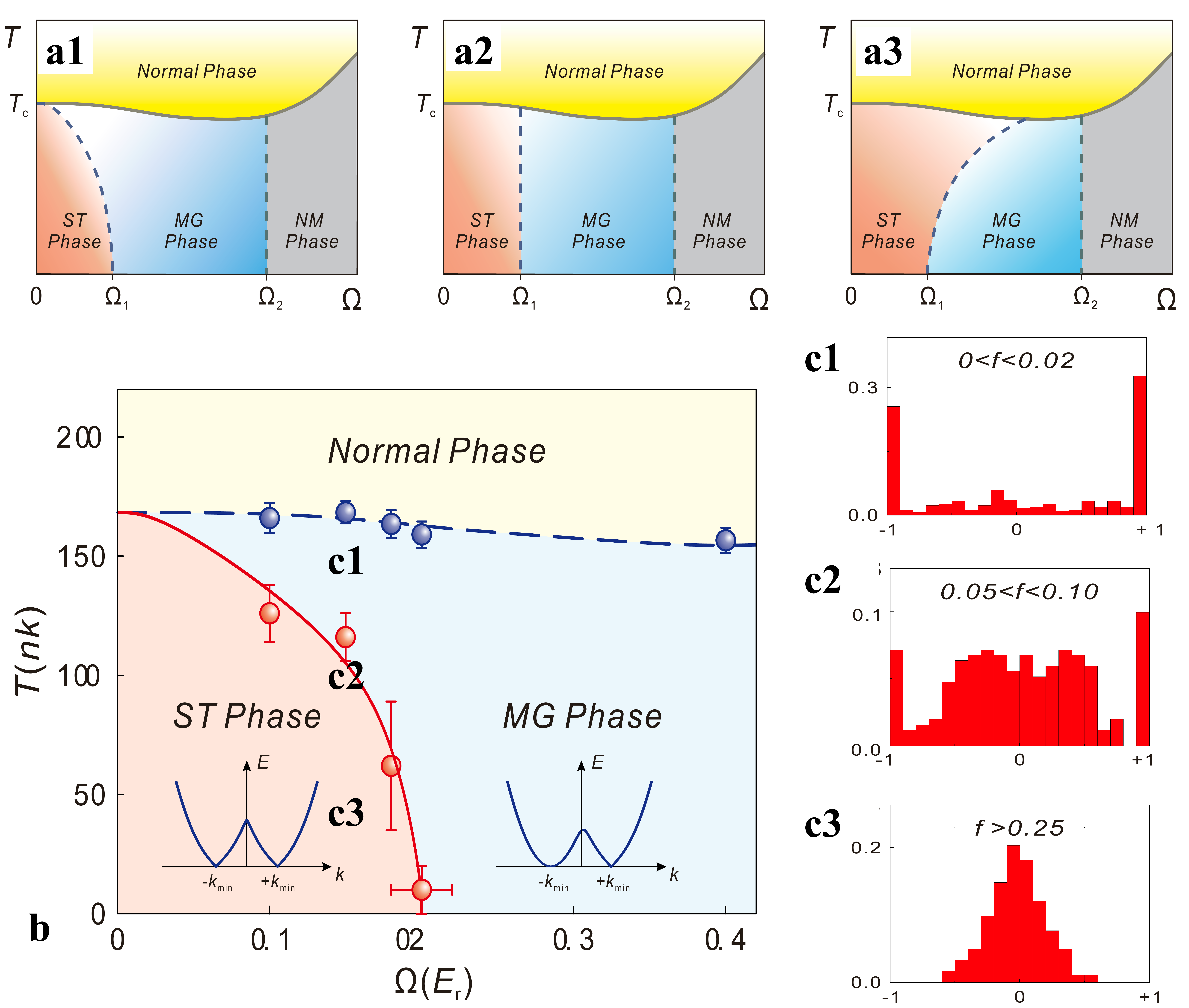}
\caption{(a) Three possible scenarios of the finite temperature phase diagram; (b) The finite temperature phase diagram determined by experimental measurements and (c) magnetization histogram for points labelled by (c1-c3) in (b). $f$ is the condensate fraction. All figures are reprinted from Ref. \cite{Shuai_FiniteT}.\label{Raman_FiniteT}}
\end{figure}

Experimentally, at the lowest temperature, a miscible to immiscrible phase transition has been observed around $\Omega=0.2 E_\text{r}$ \cite{Spielman_SOC}. However, since the period of density stripe is about $1/k_0$, which is of the order of laser wave length. Therefore it requires high resolution in-situ image to observe the density stripe order and so far has not been achieved by any group yet. Across $\Omega_2$, it has also been observed that the condensate momentum will change from zero to finite \cite{Spielman_SOC}. Moreover, since a magnetic phase transition takes place simultaneously at $\Omega_2$ where the magnetization changes from zero to finite, there should also be signatures of this magnetic phase transition. In fact, a divergent spin susceptibility has been observed through collective dipole oscillation \cite{Shuai_mode}, as we will discuss in next subsection. 

The rich phases at zero-temperature lead to many intriguing issues. One immediate generalization is what happen to phase boundary at finite temperature. Three possible scenarios of finite temperature phase diagram is schematically shown in Fig. \ref{Raman_FiniteT}(a), that is, as temperature increases, stripe phase turns into magnetized phase before entering normal state (a1); or stripe phase directly transits to normal state (a2); or magnetized phase can turn into stripe phase before entering normal state (a3). 

Magnetization histogram measurements have been proposed and carried out to determine the phase diagram experimentally \cite{Shuai_FiniteT}. It is performed as follows: for a fix parameter of SO coupling and temperature, one repeats the experiment hundreds times, and for each experiment the magnetization $M=(N_{0\uparrow}-N_{0\downarrow})/(N_{0\uparrow}+N_{0\downarrow})$ is recorded. Then one can plot the occurrence of different value of $M$. For the magnetized phase, since condensate either appears in $\varphi({\bf p}_{+})$ or appears in $\varphi({\bf p}_{-})$, and because of spin-momentum locking, these two states have opposite magnetization. Therefore, one expects two peaks at opposite $M$ for the histogram of a magnetized state. Whereas for the stripe phase, expanding the many-body wave function Eq. \ref{WF_stripe} yields a superposition of all $\psi_\text{L}^{\dag m}\psi_{\text{R}}^{\dag (N-m)}|0\rangle$ terms with $m=0,\dots,N$, which gives rise to a Gaussian distribution centered at $M=0$. In fact, as demonstrated in Ref. \cite{Shuai_FiniteT}, when condensate first forms, the magnetization histogram always appears as double peaks, as shown in Fig. \ref{Raman_FiniteT}(c1), which is consistent with the feature of a magnetized phase with $Z_2$ symmetry breaking. By further lowering temperature, the double peaks decreases and meanwhile a broad peak centered at $M=0$ starts to appear, as Fig. \ref{Raman_FiniteT}(c2). Finally,  at the lowest temperature, the double peaks completely disappear and only the peak at $M=0$ remains, as shown in Fig. \ref{Raman_FiniteT}(c), which is consistent with the feature of a stripe phase. This reveals a phase transition from magnetized phase to stripe phase with lowering temperature. By repeating such measurement for a set of different Raman coupling, a finite temperature phase diagram is experimentally constructed \cite{Shuai_FiniteT}, as shown in Fig. \ref{Raman_FiniteT}(b). A physical argument to explain this result is presented based on different behavior of low-energy excitations of these two phases \cite{Shuai_FiniteT}, as discussed in Sec. \ref{Excitation}, and a quantitative theoretical study is still absent yet.

\subsubsection{Excitations \label{Excitation}}

In this part we shall discuss three types of elementary excitations of this type of SO coupled condensate: i) collective modes in a harmonic trap; ii) phonons and roton; and iii) topological excitations of vortex and soliton. 

\vspace{0.1in}

$\circ$ \textbf{Collective modes in a harmonic trap:} Dipole oscillation is the most simplest collective mode for a BEC inside a harmonic trap. It represents a center-of-mass oscillation of all atoms. Because of the Kohn theorem, the frequency of this mode should be exactly the same as trapping frequency if the system possesses Galilean invariance. Hence, the derivation of dipole oscillation frequency from trap frequency reveals the absence of Galilean invariance of this system. Moreover, as velocity changes as a function of time during dipole oscillation, the spin of atoms also keep varying due to spin-momentum locking. Thus, spin-momentum locking also allows one to probe spin physics through dipole oscillation, which is another unique manifestation of SO coupling.

In fact, one can utilize the gauge field itself to excite dipole oscillation in this system. By suddenly changing $\delta$, the gauge field changes time dependently which generates a pulse of synthetic electric field \cite{NIST_electric}. This pulse of synthetic electric field gives a finite velocity to condensate so that the condensate starts to oscillate in the trap. It has been found that the frequency is indeed deviated from trap frequency, and to a very good approximation, the new frequency is consistent with an effective mass approximation of SO-coupling modified single particle dispersion \cite{Shuai_mode,NIST_electric}.  

\begin{figure}[t]
\includegraphics[width=3.4 in]
{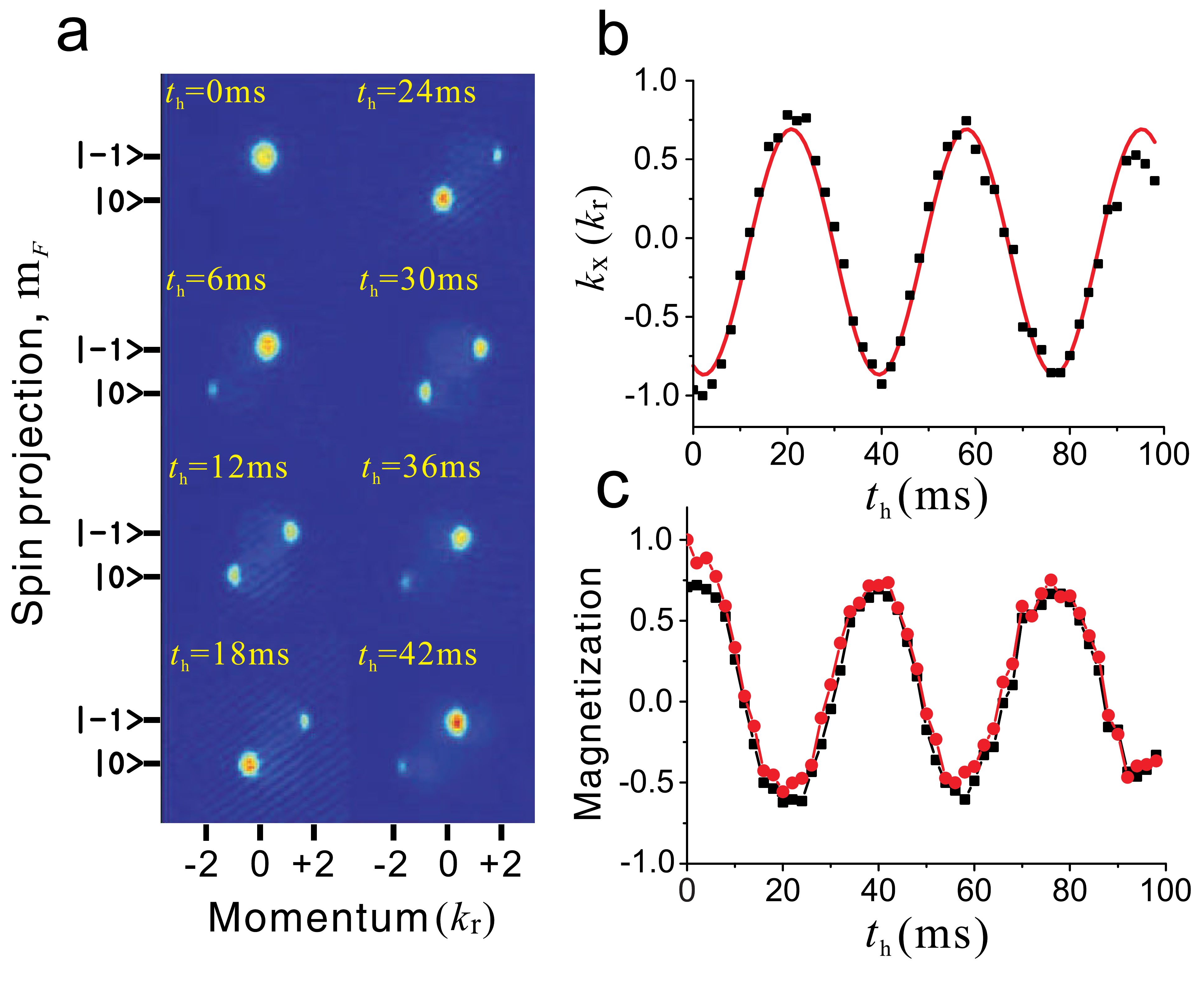}
\caption{Magnetization oscillation for $\delta=1E_\text{r}$ and $\Omega=4.8E_\text{r}$. (a) Spin-resolved TOF images for various holding times $t_\text{h}$. (b) Quasimomentum $k_x$ versus $t_\text{h}$. (c) Polarization $M$ versus $t_\text{h}$. The red circles are directly measured, while the black squares are deduced from (b) (see Ref. \cite{Shuai_mode} for details). Reprinted from Ref. \cite{Shuai_mode}. \label{Spin_oscillation}}
\end{figure}

\begin{figure}[t]
\includegraphics[width=3.0 in]
{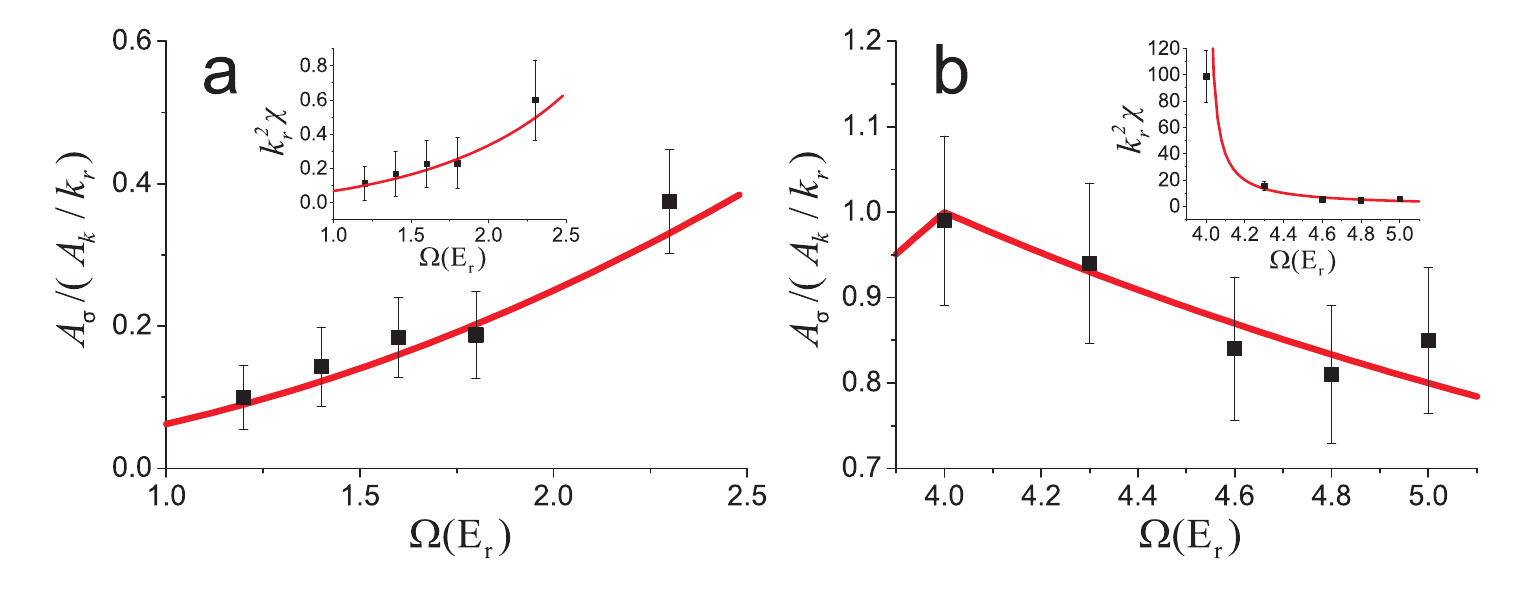}
\caption{Amplitude ratio of spin and momentum oscillation $A_{\sigma}/(A_{k}/k_{\text{r}})$ versus $\Omega$, 
  for magnetic phase (a) and nonmagnetic phase (b).
  The inset is for the spin polarization susceptibility $E_{\text{r}} \, \chi$ deduced from this ratio
  via Eq. \ref{relation}. The red solid line is from the solution of Eq. \ref{chi}. Reprinted from Ref. \cite{Shuai_mode}. \label{Spin_polarization}}
\end{figure}

Moreover, by simultaneously measuring condensate momentum through time-of-flight and spin population through Stern-Gerlach, one can find that a spin oscillation is induced by dipole oscillation and they are well synchronized \cite{Shuai_mode}, as shown in Fig. \ref{Spin_oscillation}.
As discussed in sub-Section \ref{Single}, in the moving frame, the Hamiltonian acquires a velocity dependent $v_xk_0\sigma_z/m$ term. Thus, for small amplitude dipole oscillation, it is equivalent to applying an ac-Zeeman field to the system. It is therefore easy to understand that dipole oscillation is connected to spin susceptibility. The relation between dipole oscillation and spin susceptibility is more rigorously established by the sum-rule approach \cite{Stringari_sumrule}. Ref. \cite{Stringari_sumrule} shows that a lower bound of dipole oscillation frequency can be obtained by optimizing following equation with respect to $\eta$
\begin{equation}
\omega^2_\text{d}=\frac{m_{1}(F)}{m_{-1}(F)}=\frac{-2\eta^2 k_0^2\Omega\langle\sigma_x\rangle+\omega^2_x}{1+(1+\eta)^2k_0^2\chi}.
\end{equation}   
$F=\sum_{i=1}^{N}(k_{x,i}+\eta k_0\sigma_{z,i})$ is a linear combination of spin and momentum, and such a combination reflects the spin and momentum are coupled. $\chi$ is the spin susceptibility. Ignoring the difference between $g$ and $g_{\uparrow\downarrow}$, $\chi$ is given by
\begin{equation}
  \chi = \left\{ 
         \begin{array}{lcl} 
   \frac{1}{2E_\text{r}}        \frac{(\Omega/\Omega_2)^2}{1-(\Omega/\Omega_2)^2} \hspace{5mm} & \mbox{for} & \Omega < \Omega_2 \\
   \frac{1}{2E_\text{r}}  	   \frac{1}{ \Omega/\Omega_2-1}              \hspace{5mm} & \mbox{for} & \Omega > \Omega_2
	  \end{array} 
	  \right.
\label{chi}
\end{equation} 
As discussed in the previous Section, at $\Omega=\Omega_2$ the system undergoes a transition from nonmagnetic phase to magnetic field, thus, $\chi$ diverges at $\Omega=\Omega_2$. Ref. \cite{Stringari_sumrule} also shows that $\chi$ can be deduced from the oscillation amplitudes as
\begin{equation}
\frac{A_\sigma}{A_k/k_{\text{0}}}=\frac{E_{\text{r}}\chi}{1+E_{\text{r}}\chi} \; . \label{relation}
\end{equation}
where $A_\sigma$ and $A_k$ are spin and momentum oscillation amplitudes, respectively. Eq. \ref{chi} and Eq. \ref{relation} are verified experimentally \cite{Shuai_mode}, as shown in Fig. \ref{Spin_polarization}, from which a divergent spin susceptibility is identified as a signature of the magnetic phase transition. Dipole oscillation has also been studied by a variational wave function approach, in which the coupling between dipole mode and other breathing modes are also discussed \cite{Chen_dipole}.

\vspace{0.1in}

\begin{figure}[t] 
\includegraphics[width=3.0 in]
{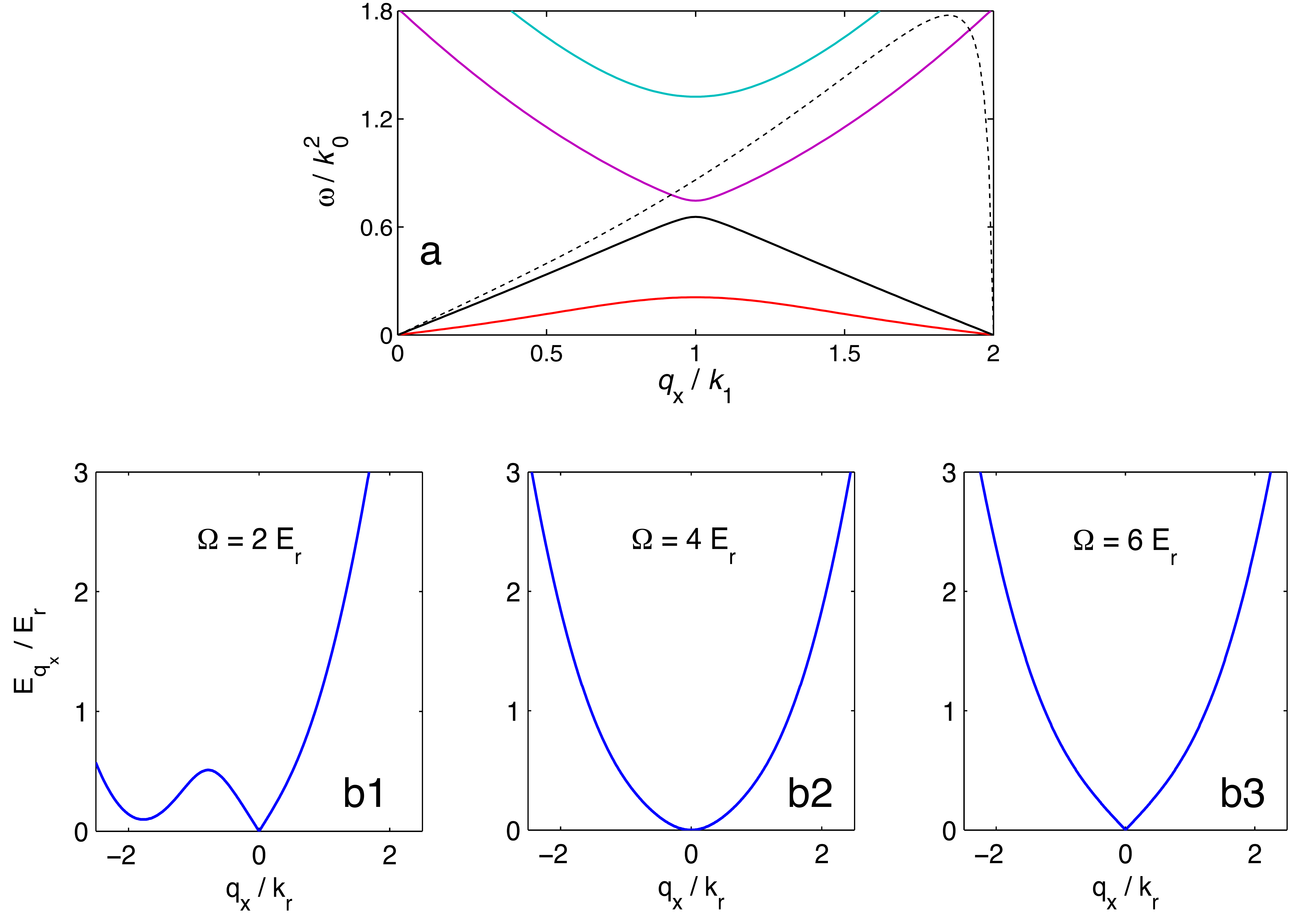}
\caption{(a) Bogoliubov modes of stripe phase, reprinted from Ref. \cite{Superstripe}; (b1-b3) Bogoliubov modes for plane wave phase, reprinted from Ref. \cite{Zheng_Yu}. See Ref. \cite{Superstripe} and Ref. \cite{Zheng_Yu} for details.  \label{Bogoliubov}}
\end{figure}

$\circ$ \textbf{Phonon and roton:} Here we discuss low-energy excitations of a uniform system. The main features are highlighted in Fig. \ref{Bogoliubov}.

\vspace{0.05in}

i) In the stripe phase, excitation spectrum exhibits two linear phonon modes and band structure \cite{Superstripe}, as shown in Fig. \ref{Bogoliubov}(a). This is a manifestation of translational symmetry breaking. One of the linear phonon mode corresponds to breaking $U(1)$ gauge symmetry and another linear mode corresponds to breaking of spatial translational symmetry.

\vspace{0.05in}

ii) In the plane wave phase when $\Omega<\Omega_2$, the system breaks $Z_2$ symmetry. If bosons condense in $\varphi({\bf p}_+)$ state, it will exhibit a linear mode nearby ${\bf p}_+$. On the other hand, the excitation spectrum exhibits a local minimum around ${\bf p}_{-}$, as shown in Fig. \ref{Bogoliubov}(b1) \cite{Stamper-Kurn,Zheng_Yu,roton}. This indicates, though the system does not form a density stripe order, it still has a tendency toward a density wave order phase. Such a minimum is reminiscent of roton structure in the excitation spectrum of superfluid Helium, which is an indication that Helium liquid has a tendency toward crystallization. 

Therefore, at the critical regime of $\Omega=\Omega_1$, the stripe phase displays two linear modes at the lowest energy, whereas the plane wave phase displays a linear mode and a quadratic minimum with vanishingly small roton gap at the lowest energy. Comparing these two cases, the plane wave phase has larger low-energy density-of-state, and therefore it gains more entropy at finite temperature. This explains the finite temperature phase diagram reported in Ref. \cite{Shuai_FiniteT} where the plane wave phase is more favorable as temperature increases. This explanation connects symmetry breaking properties with thermodynamics. 

\vspace{0.05in}

iii) At $\Omega=\Omega_2$, single-particle spectrum exhibits a $k^4_x$ dispersion, and consequently, its Bogoliubov spectrum display a $k^2_x$ behavior at this point, as shown in Fig. \ref{Bogoliubov}(b2) \cite{Zheng_Yu,roton}. The sound velocity vanishes at this point. Finally, when $\Omega>\Omega_2$, the excitation spectrum becomes the same as a normal superfluid, as shown in Fig. \ref{Bogoliubov}(b3).

\vspace{0.05in}

\begin{figure}[tbh] 
\includegraphics[width=3.3 in]
{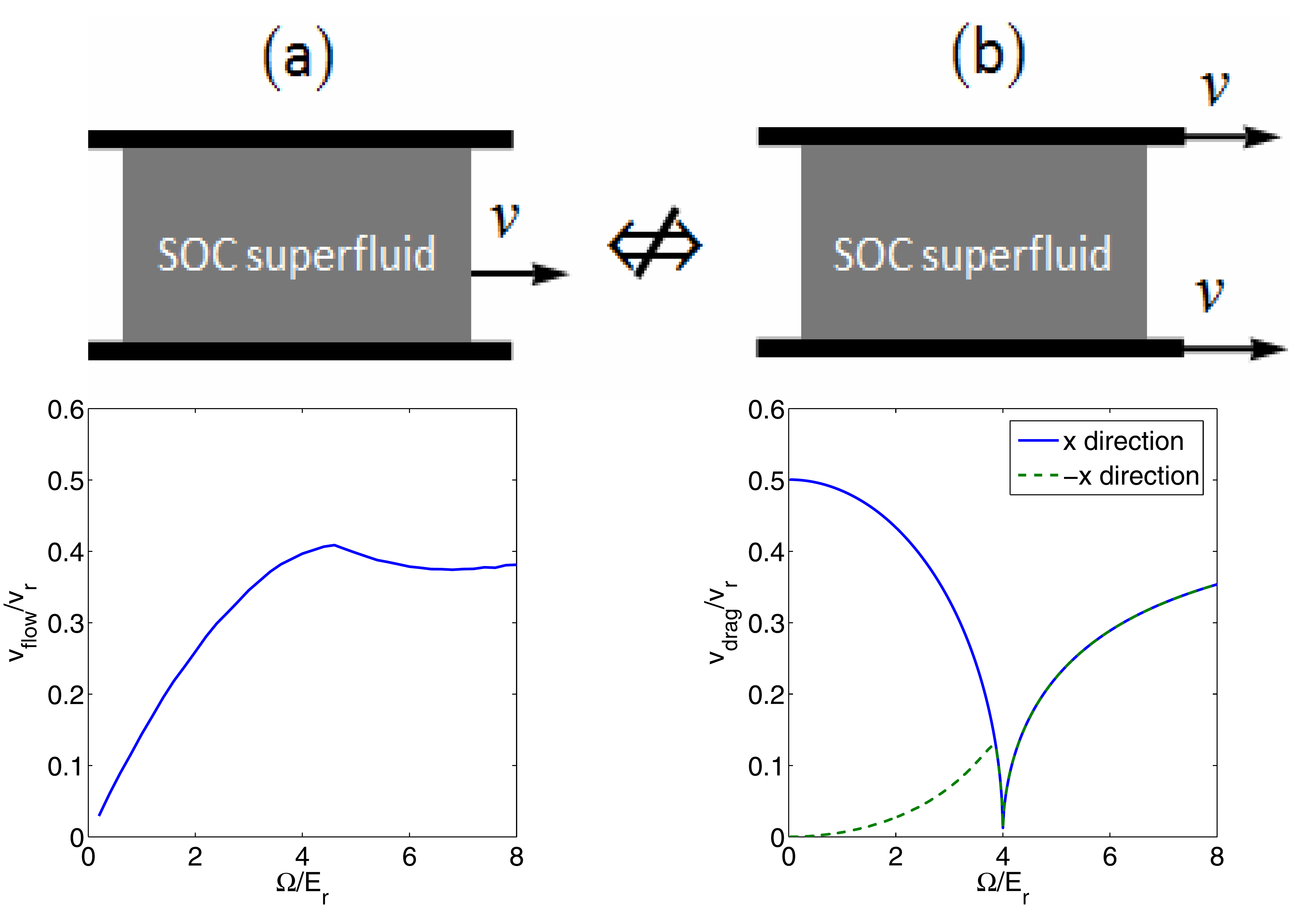}
\caption{(a) Critical flowing velocity for a condensate moving with constant velocity $v$ against static impurities; (b) Critical dragging velocity for impurity moving with a velocity $v$ inside a static condensate. Reprinted from Ref. \cite{Wu_Biao} and Ref. \cite{Zheng_Yu}.   \label{critical}}
\end{figure}

In a system with Galilean invariance, the superfluid critical velocity is always given by $v_\text{c}=\min(\epsilon(p)/p)$. If that is the case, the superfluid critical velocity will vanish when the phonon mode has a quadratic dispersion.  Nevertheless, the absence of Galilean invariance gives rise to a richer behavior of superfluid critical velocity. Fig. \ref{critical} illustrates two types of measurements of critical velocity. These two types of measurements yield different results, as first pointed out in Ref. \cite{Wu_Biao} using Rashba SO coupling as an example. 

Fig. \ref{critical}(a) describes the case that a Bose condensate moves with a constant velocity $v$ against static impurities. In this case, it probes when the excitation of a moving condensate becomes unstable in the laboratory frame. As discussed in Sec. \ref{Single}, a moving SO-coupled system acquires an additional $v_xk_0\sigma_z/m$ term, with which the single particle dispersion no longer displays a $k_x^4$ behavior. Therefore the critical velocity from this type of measurement is always finite \cite{Zheng_Yu,dynamic_instability}. Fig. \ref{critical}(b) shows the second type of measurement that corresponds to impurities moving with a constant velocity $v$ inside a static condensate. This probes the spectrum of a static condensate and obeys Landau's argument, which is given by  $v_\text{c}=\min(\epsilon(p)/p)$. In this case, the critical velocity vanishes at $\Omega=\Omega_2$, as shown in Fig. \ref{critical}(b) \cite{Zheng_Yu,dynamic_instability}.

\vspace{0.1in}

$\circ$ \textbf{Vortex and soliton:} Vortex is the most fundamental topological object in a superfluid. Usually vortex can be created by rotating a condensate. Rotating a SO coupled condensate becomes quite nontrivial because angular momentum is no longer a good quantum number due to SO coupling. Ref. \cite{Galitski_vortex} discussed various schemes to rotate a condensate in presence of SO coupling. Moreover, as a vortex generates a local velocity current winding around the center of vortex core, because of the spin-momentum locking, the local velocity current will also lead to a spin density distribution around the vortex core, as shown in Ref. \cite{Galitski_vortex}. Vortex dynamics in a non-rotating condensate has also been studied by Ref. \cite{Fetter}.

\begin{figure}[th] 
\includegraphics[width=3.3 in]
{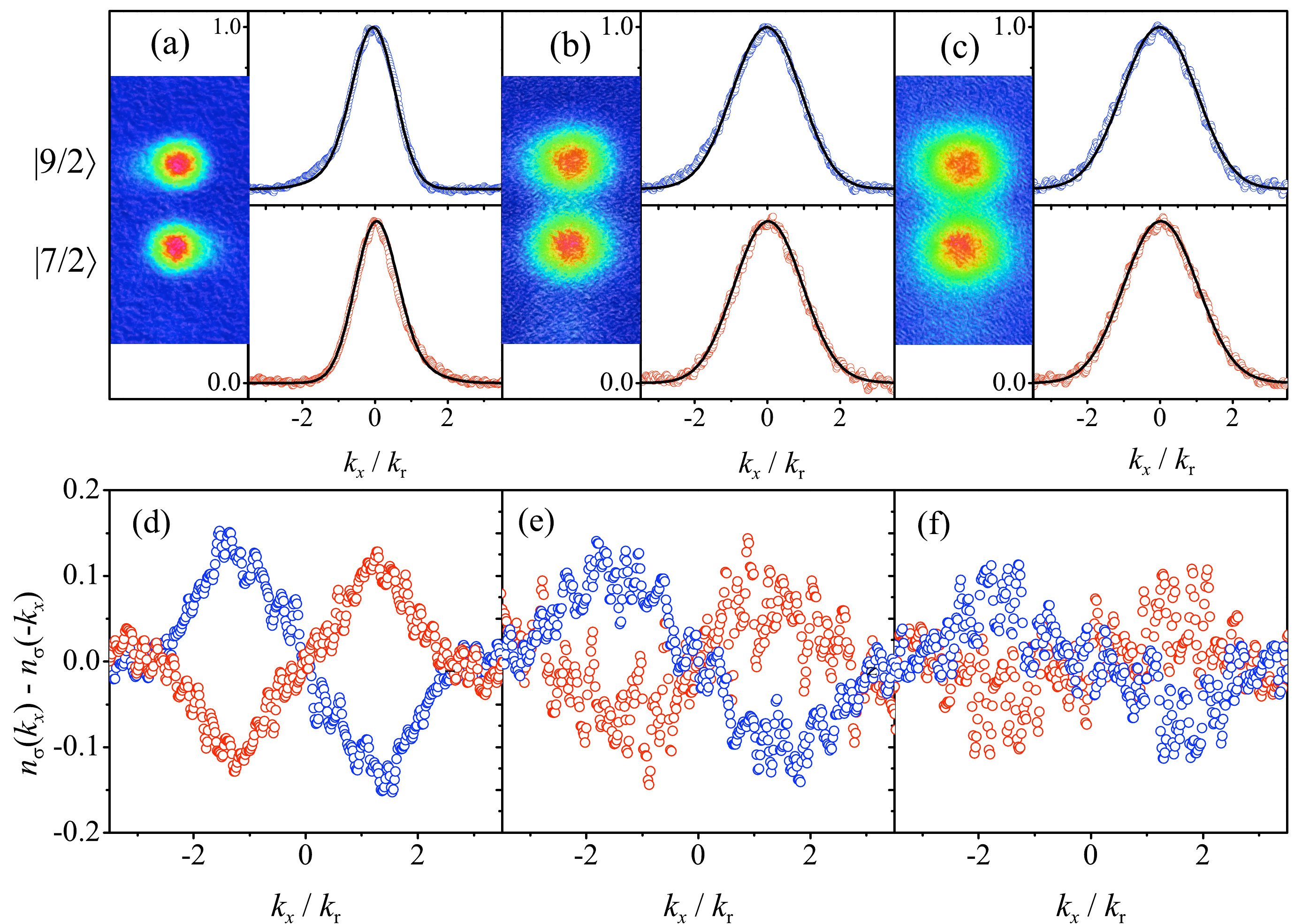}
\caption{(a-c) Momentum distribution $n_\sigma(k_x)$ for two different spin components; (d-f) $n_\sigma(k_x)-n_\sigma(-k_x)$ for different spin components. Reprinted from Ref. \cite{Jing_SOC}. For more details, see Ref. \cite{Jing_SOC}. \label{momentum_Fermion}}
\end{figure}

Another widely studied topological object in a superfluid is soliton. Usually a bright (dark) soliton has a smooth sech$^2$-shape density (density dip) distribution, and in a scalar condensate such a soliton keeps its density distribution as it travels. It has been found that both dark and bright can be changed significantly by SO coupling \cite{Fialko,Wu_soliton,Pelinovshy,Schmelcher}. When the single particle dispersion displays double minimum, there exists two degenerate soliton solutions whose momentum is located nearby each local minimum. Their superposition gives rise to soliton with spatially density modulation, in contrast to sech$^2$ smooth density distribution of conventional solitons  \cite{Wu_soliton,Pelinovshy,Schmelcher}. This is reminiscent of ground state stripe phase of SO coupled condensate, and is therefore named as ``stripe soliton". In addition, because the lack of Galilean invariance, when soliton moves with a finite velocity, an additional velocity-dependent Zeeman field removes the degeneracy in the single particle dispersion. Thus, at large enough velocity, the density modulation of soliton gradually disappears and its density distribution changes back to a smooth form \cite{Wu_soliton}. Another type of gap soliton has also been studied in SO coupled condensate by assuming the spatially periodic coupling term $\Omega(x)$ \cite{Abdullaev}.  
 
\subsubsection{Dynamics}

SO-coupling-modified single particle spectrum exhibits gapped Dirac point and spin-momentum locking, which can also give rise to rich quantum dynamics. In this section we summarize some recent experimental progresses of quantum dynamics in SO coupled condensates. 

\vspace{0.1in}

$\circ$ \textbf{Landau-Zener Tunneling and Zitterbewegung:} The single particle dispersion with SO coupling exhibits a Dirac point-like dispersion, although the Dirac point is gapped by finite Raman coupling $\Omega$-term. If one smoothly changes the momentum of atoms to cross the Dirac point, the system will exhibit a Landau-Zener tunneling in which the transition probability depends on how fast the parameter changes as it crosses the Dirac point. This has been experimentally studied by Ref. \cite{Yong_LZ}. If one fast quenches the parameter nearby the Dirac point, it will induce Zitterbewegung. Zitterbewegung describes a trembling motion accompanied by spin oscillation, which is first predicted for relativistic Dirac equation but it is difficult to be observed for electrons. An analogy of Zetterbewegung has only been observed recently in a trapped ion system. In our system, the oscillation frequency of Zitterbewegung is controlled by the gap of Dirac point. This phenomenon has been experimentally observed in SO coupled condensate in Ref. \cite{Spielman_ZB,Engels_ZB}.

\vspace{0.1in}

$\circ$ \textbf{Classical Spin-Hall Effect:} It has been proposed that a class spin-Hall effect can be induced by an effective spin-dependent gauge field \cite{Zhu_SHE,Liu_SHE} and can be detected through dipole mode \cite{Duine}. In this Raman-induced SO system, since the minimum of single-particle dispersion is located at finite momentum $k_{\pm}$, and due to spin-momentum locking, nearby $k_{\pm}$ the spin polarization is nearly opposite, the low-energy physics can be approximately described by $(p_x-A_x\sigma_z)^2/(2m)$. $A_x$ can be controlled by Raman coupling strength $\Omega$. In experiment Ref. \cite{Spielman_SH}, they let $\Omega$ depend on $y$, and therefore $A_x$ depends on $y$. This gives rise to a magnetic field ${\bf B}=\partial A_x/\partial y \hat{z}\sigma_z$ that is opposite for different spin component. This spin-dependent Lorentz force leads to classical spin-Hall effect as observed in Ref. \cite{Spielman_SH}, that is, when one drives a particle current along $\hat{x}$, spin-up atoms will acquire a momentum along $\hat{y}$ while spin-down atoms will acquire a momentum along $-\hat{y}$. 

\begin{figure}[tb] 
\includegraphics[width=3.3 in]
{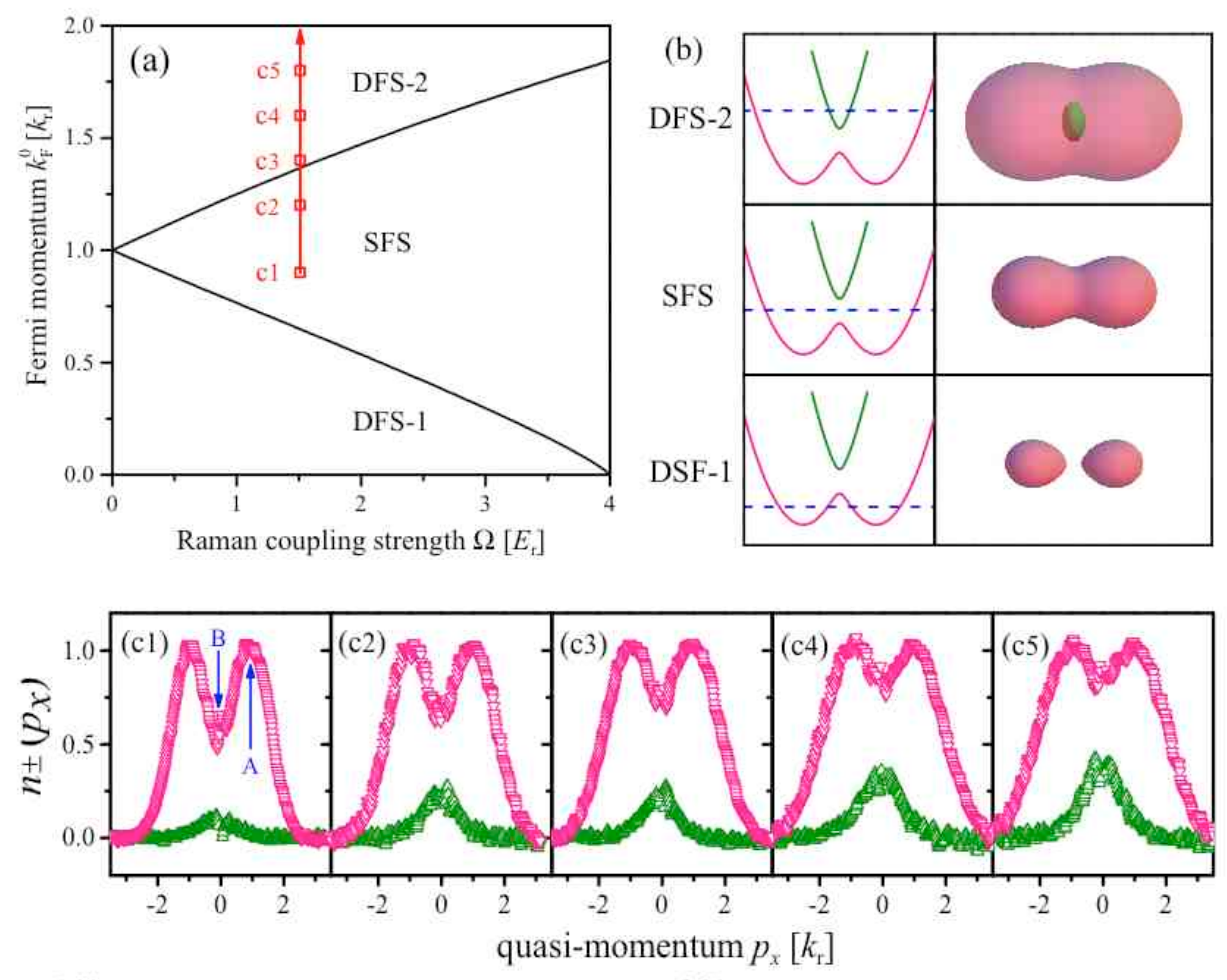}
\caption{(a) Theoretical phase diagram at $T=0$.
$k^0_{\text{F}}=\hbar(3\pi^2 n)^{1/3}$. ``SFS" means
single Fermi surface. ``DFS" means double Fermi surface. (b) Illustration of different topology of Fermi surfaces. The single particle
energy dispersion is drawn for small $\Omega$. Dashed blue line is the chemical potential.  (c) Quasi-momentum
distribution in the helicity bases. Red and green points are
distributions for $s=-1$ and $s=1$ helicity branches, respectively. Parameters for (c1-c5) are marked in (a).
 Reprinted from Ref. \cite{Jing_SOC}. For more details, see Ref. \cite{Jing_SOC}. \label{Lifshitz}}
\end{figure}

\subsection{Fermions}

\subsubsection{Free Fermions}

Applying the same Raman coupling to ${}^{40}$K or ${}^6$Li, experimentally SO coupling effect has also been realized for a degenerate Fermi gas \cite{Jing_SOC,MIT_SOC}. Even for noninteracting fermions, a fermion system displays distinct phenomena compared to the boson system, because fermions populate all momentum states below the Fermi energy and therefore more information can be revealed about the SO-coupling-modified dispersion. First, because spin-momentum locking, when spin-resolved momentum distribution is measured by combing time-of-flight with Stern-Gerlach method, one can clearly visualize the difference between $n_\uparrow(k_x)$ and $n_\downarrow(k_x)$, and the difference between $n_\sigma(k_x)$ and $n_\sigma(-k_x)$ for each spin component because of the parity symmetry breaking, as shown in Fig. \ref{momentum_Fermion}.

Secondly, with modified dispersion, the topological structure of the Fermi surface will exhibit a series of changes as the Fermi energy increases. As shown in Fig. \ref{Lifshitz}, for the lowest density, only the helicity $s=-$ branch $E_{{\bf k},-}$ is occupied and the Fermi surfaces are two disjointed spheres, as shown in Fig. \ref{Lifshitz}. As density increases, these two disjointed Fermi surfaces gradually merge into a single Fermi surface. Later as density further increases, the helicity $s=+$ branch also becomes occupied and there will be a new Fermi sphere emerging from the center of the larger Fermi surface. Such a series of changes can be revealed from occupation of $s=\pm$ branches as shown in Fig. \ref{Lifshitz}(c).

\begin{figure}[tb] 
\includegraphics[width=3.3 in]
{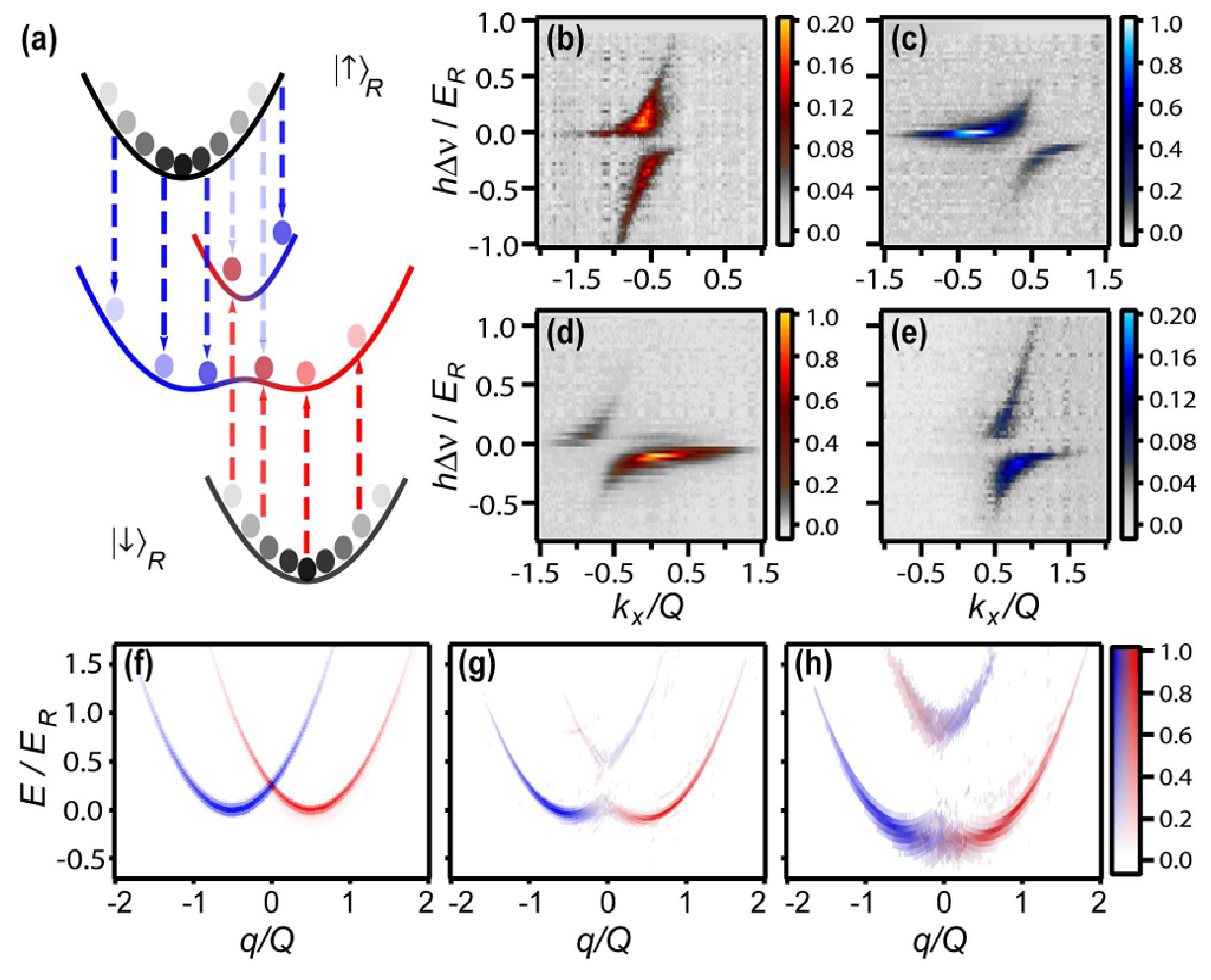}
\caption{(a) Schematic of spin injection spectroscopy. (b-e) Spin-resolved dispersion from spin injection spectroscopy. (f-h) Reconstructed spin-resolved dispersion. 
 Reprinted from Ref. \cite{MIT_SOC}. For more details, see Ref. \cite{MIT_SOC}. \label{spin_injection}}
\end{figure}

Comparing to momentum distribution, spectroscopy is a more powerful tool to reveal single particle band properties. Ref. \cite{MIT_SOC} reported spin injection spectroscopy. By controlling frequency of radio-frequency waves, atoms initially prepared in a third state can be selectively transmitted to either $\uparrow$ or $\downarrow$ state with same momentum, as shown in Fig. \ref{spin_injection}(a). This method can not only measure the dispersion, but also reveal the spin component of each momentum eigenstate. The typical experimental results are shown in Fig. \ref{spin_injection}(b-e) and the reconstructed spin-resolved dispersions are shown in Fig. \ref{spin_injection}(f-h). The locking between spin and momentum are proved clearly in this measurement. 

\begin{figure}[tb] 
\includegraphics[width=3.3 in]
{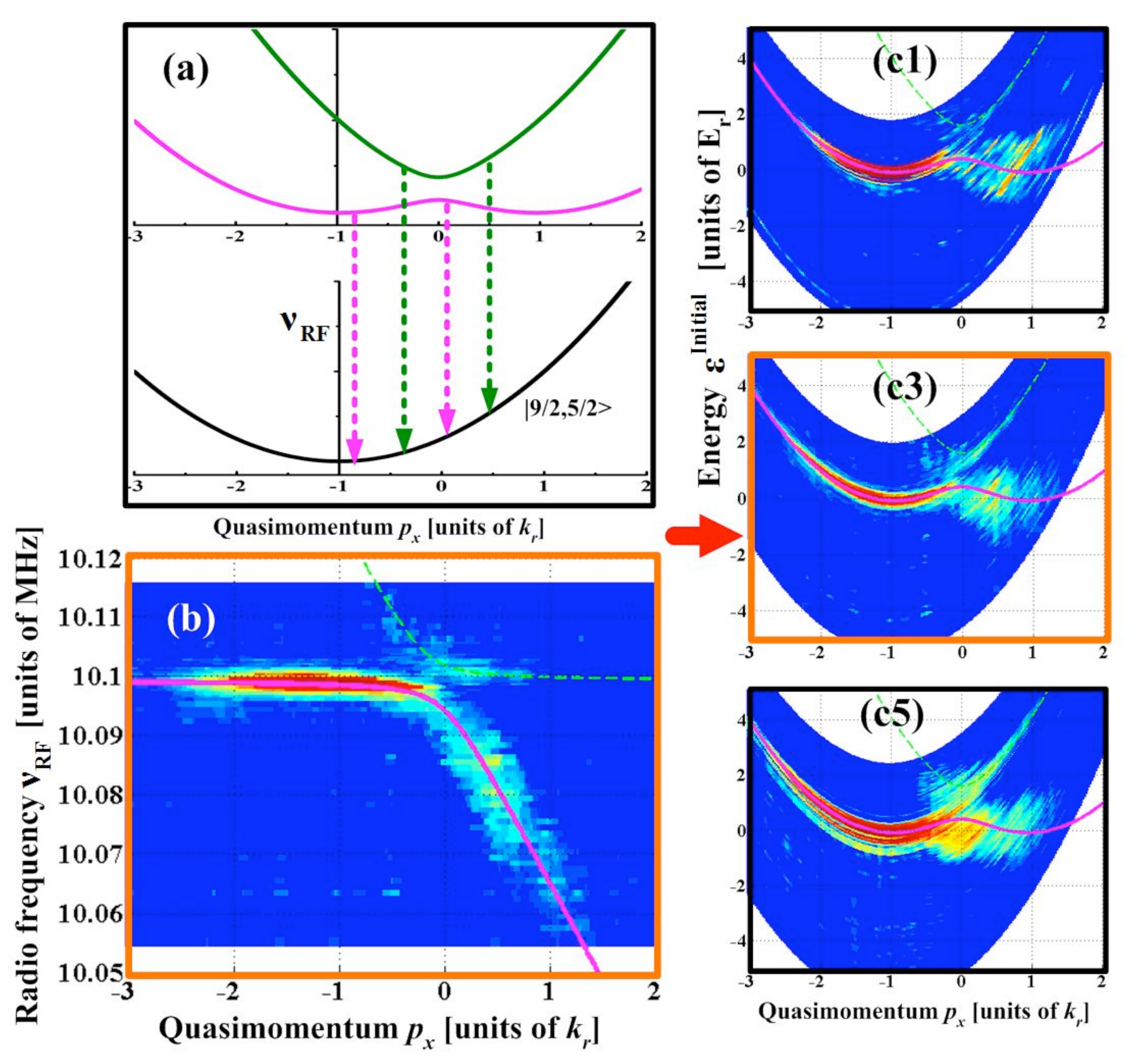}
\caption{(a) Schematic of momentum-resolved radio-frequency spectroscopy of SO coupled Fermi gases. (b)  Intensity map of the atoms in $|9/2,5/2\rangle$ state as a function of
($\nu_{\rm RF}, k_{x}$) plane. (c) Single particle dispersion and atom population measured for (c1), (c3) and (c5) in Fig. \ref{Lifshitz}.
 Reprinted from Ref. \cite{Jing_SOC}. For more details, see Ref. \cite{Jing_SOC}. 
 \label{rf_spectroscopy}}
\end{figure}

Ref. \cite{Jing_SOC} reported momentum-resolved radio-frequency spectroscopy. This is somewhat opposite to spin injection spectroscopy. Atoms are initially prepared in a system with SO coupling, and then radio-frequency is applied to drive transition from one of the spin states to a third state, as schematically shown in Fig. \ref{rf_spectroscopy}. This method can not only determine the spin-resolved dispersion, as shown in Fig. \ref{rf_spectroscopy}(b), but also reveal the change of Fermi sea, since only the occupied initial states can undergo such a transition to the third state. As shown in Fig. \ref{rf_spectroscopy}(c1-c3), at lower density, fermions only populate $s=-$ band, and as density increases, fermions start to populate both $s=-$ and $s=+$ bands.

Finally, SO coupling also plays an important role in spin diffusion, for instance in semiconductors. Such an effect has also been observed in cold atom experiment \cite{Jing_SOC}. Starting with a fully polarized Fermi sea, fermion with different momentum undergoes spin procession in different way, and soon the total spin oscillation will be damped, as observed in Ref. \cite{Jing_SOC}. Recently, a number of theoretical papers have also studied spin diffusion in SO coupled Fermi gases \cite{dynamics_1,dynamics_2,dynamics_3,dynamics_4}, whose predictions can be verified in future experiments. 

\subsubsection{Feshbach Resonance}

\begin{figure}[tb] 
\includegraphics[width=2.5 in]
{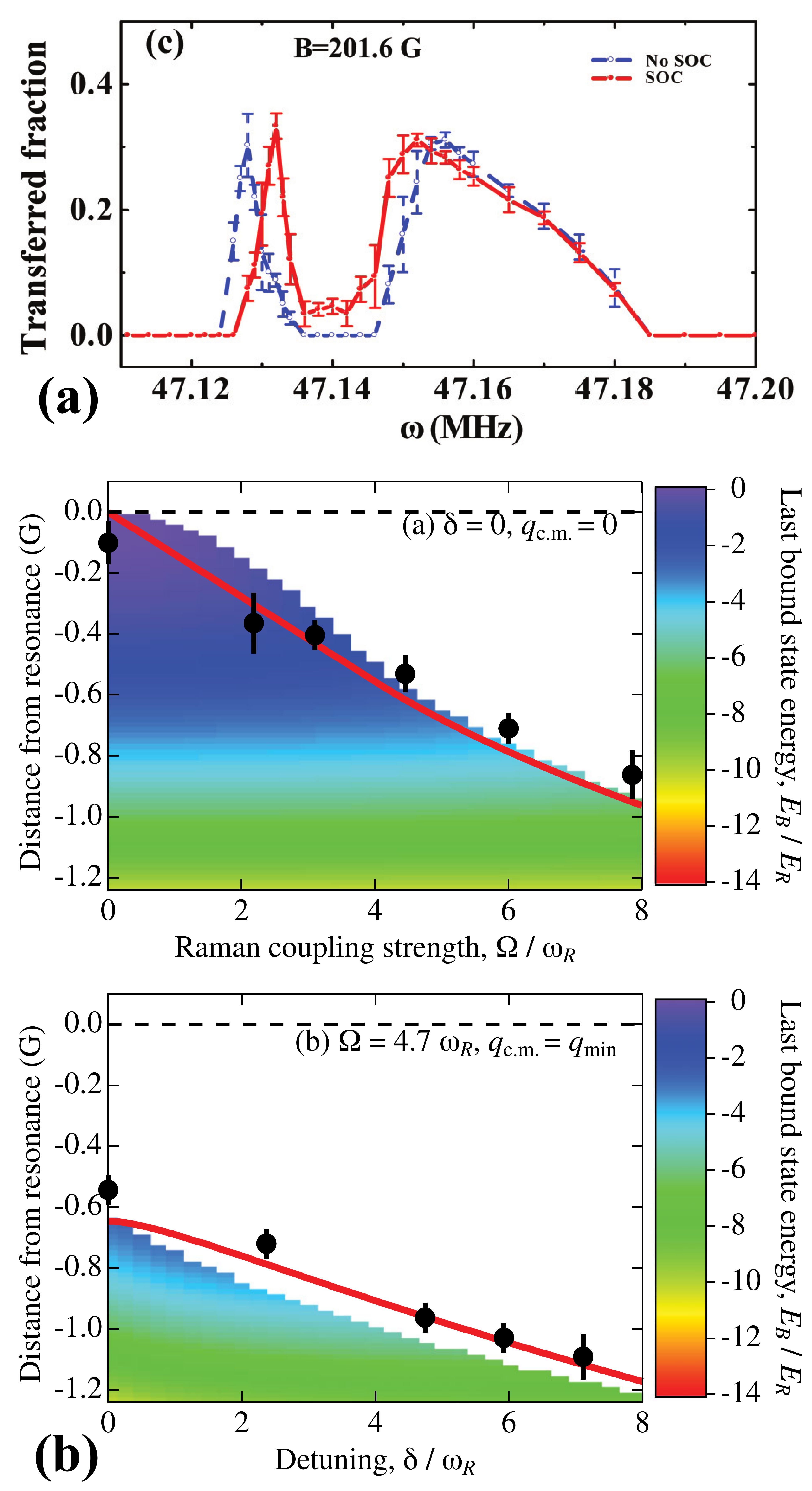}
\caption{(a) The integrated radio-frequency spectroscopy for $^{40}$K at $201.6$G with $a_\text{s}\simeq 2215.6a_\text{B}$ and $1/(k_\text{F}a_\text{s})\simeq 0.66$. The narrow and broad peaks correspond to radio-frequency response of free atoms and bound pair, respectively. Reprinted from Ref. \cite{Jing_spectroscopy}. (b) Predicated and measured shallow bound state energy as a function of $\Omega$ and $\delta$. Reprinted from Ref. \cite{Speilman_molecule}. \label{Molecule_Raman}}
\end{figure}

Nearby a Feshbach resonance, SO coupling significantly changes physics of two-body bound states:

First, in contrast to Rashba SO coupling, the Raman-induced SO coupling does not significantly enhance single particle low-energy DoS. Instead, due to the mixing of singlet and triplet channel, the weight of two-body wave function in the singlet channel is reduced, and therefore it effectively suppresses the formation of bound state. This two-body problem is calculated in Ref. \cite{Hu_spectroscopy} where radio-frequency spectroscopy on bound state is also investigated. Later, molecule binding energy has been experimentally measured \cite{Jing_spectroscopy} and compared with theory \cite{Hu_spectroscopy, Jing_spectroscopy}. As shown in Fig. \ref{Molecule_Raman}(a), the separation between free atom peak (narrow one) and bound pair peak (broad one) is suppressed when Raman-induced SO is turn on, which means the reduction of molecular binding energy. The suppression of binding energy also means that resonance will be shifted to positive $a_\text{s}$ side with SO coupling, and the shift will increase with the increasing of $\Omega$ or $\delta$ \cite{Peng_renormalization,Speilman_molecule,Vijay_molecule,Melo_molecule}. This is equivalent to say, for a given positive scattering length, resonance can be induced by changing $\Omega$ or $\delta$ \cite{Peng_renormalization}. This phenomenon has also been experimentally verified, as shown in Fig. \ref{Molecule_Raman}(b).

\begin{figure}[tb] 
\includegraphics[width=3.3 in]
{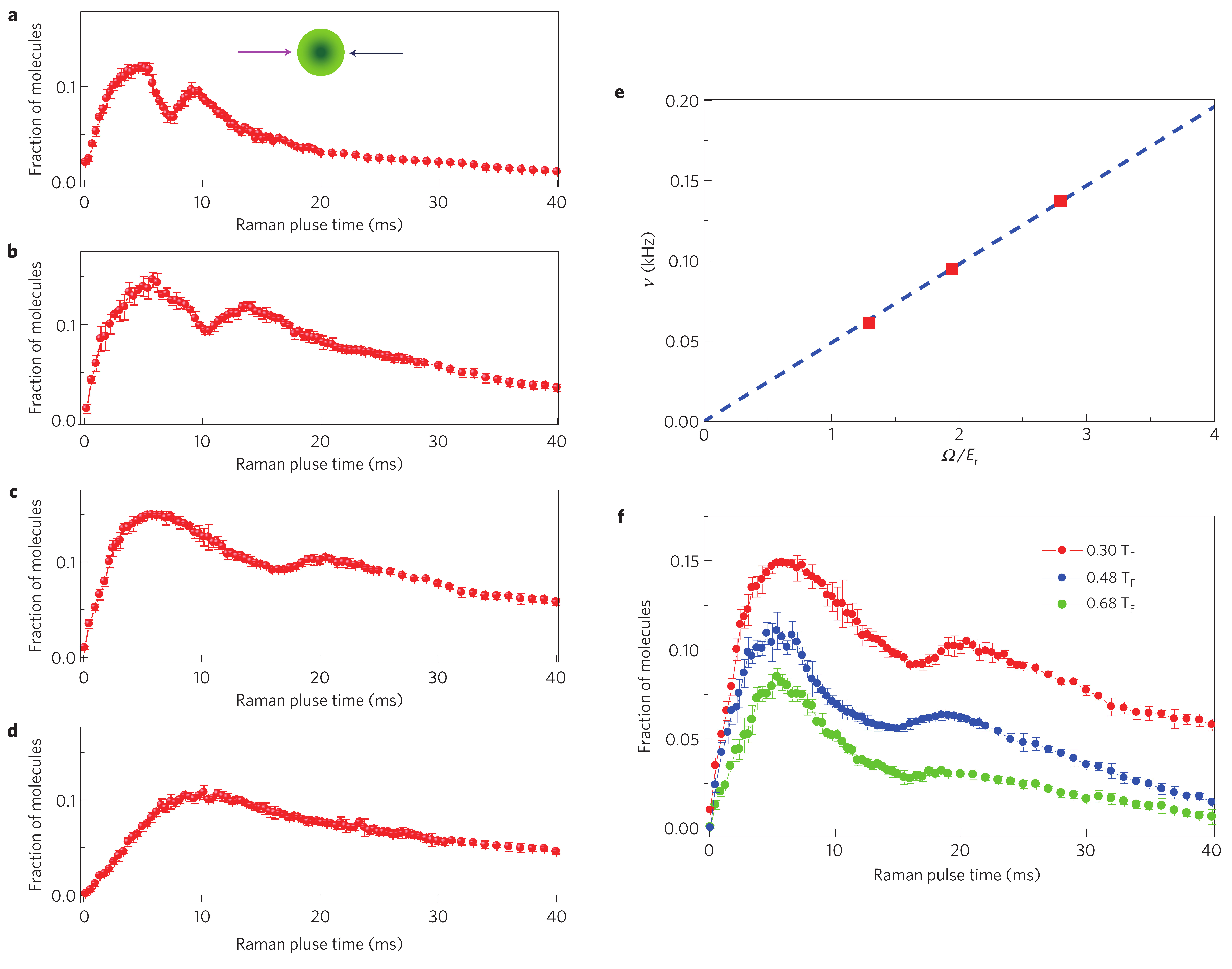}
\caption{(a-d): The dependence on the Raman
coupling strength for coherent Rabi oscillations
between a spin polarized Fermi gas and Feshbach molecular state.
Temperature $T/T_\text{F}=0.3$, and the Raman coupling strength is
$\Omega=2.8E_{\text{r}}$ (a), $\Omega=1.95E_{\text{r}}$ (b),
$\Omega=1.3E_{\text{r}}$ (c), and $\Omega=0.65E_{\text{r}}$ (d),
respectively. (e) The Rabi frequencies
obtained from (a-c) as the function of the Raman coupling strength.
(f) The dependence on the temperature  for coherent Rabi oscillations. The Raman coupling strength
$\Omega=1.3E_{r}$. Initial $T/T_{\text{F}}=0.3$ for red curve,
$T/T_{\text{F}}=0.48$ for blue curve, and $T/T_{\text{F}}=0.68$ for
green curve. Reprinted from Ref. \cite{Rabi}.\label{Rabi}}
\end{figure} 

Secondly, as discussed in Sec. \ref{TB}, SO coupling couples singlet channel to triplet channel, and the bound state wave function contains both single and triplet components. This is directly related to topological superfluid property discussed later. To experimentally verify this, one can start with a fully polarized $\left|\uparrow\right\rangle$ Fermi gas and apply a spin flip transition from $\left|\uparrow\right\rangle$ to $\left|\downarrow\right\rangle$. If this transition is momentum-independent, for instance, driven by a radio-frequency field or two parallel Raman lasers, it will not couple fully polarized gas to singlet molecules. While if this transition is momentum-dependent and possesses the effect of SO coupling, a coherent coupling between fully polarized Fermi gas and singlet molecules can be observed, as reported in Ref. \cite{Rabi}. As shown in Fig. \ref{Rabi}, the period of this many-body Rabi oscillation scales linearly with Raman coupling $\Omega$, and the period is not sensitive to temperature. This is a clear evidence of coherent Rabi oscillation. While the oscillation amplitude gradually decreases as temperature increases \cite{Rabi}. 

Thirdly, due to the absence of Galilean invariance, the binding energy varies as the center-of-mass momentum of the pair changes. Consequently, in some regime, even the bound state appears for small center-of-mass momentum, it will merge into scattering continuum for sufficient large momentum. And in some case, the minimum of binding energy will take place at finite center-of-mass momentum \cite{Vijay_molecule,Han_molecule}. This indicates that the pairs are more stable when they acquires a finite momentum, which implies a many-body FFLO type superfluid in this system.

\subsubsection{Interacting Many-body Physics}

In most cases, the interaction between two spin-$1/2$ fermions is described by
\begin{equation}
H_\text{int}=g\int d^d{\bf r}\psi^\dag_\uparrow({\bf r})\psi^\dag_\downarrow({\bf r})\psi_\downarrow({\bf r})\psi_\uparrow({\bf r}).
\end{equation}
For three-dimension, $g$ is related to scattering length $a_\text{s}$ via
\begin{equation} 
\frac{m}{4\pi a_\text{s}}=-\frac{1}{g}+\int \frac{d^3{\bf k}}{(2\pi)^3}\frac{1}{{\bf k}^2/m}. \label{renormalization}
\end{equation}
When SO coupling is introduced to atomic gases, it has been discussed in previous sec. \ref{TB} that the renormalization condition Eq. \ref{renormalization} remains unchanged.
For negative $a_\text{s}$, the system exhibits BCS instability, while for positive $a_\text{s}$, this interaction supports a two-body bound state whose energy is $\hbar^2/(ma^2_\text{s})$. Thus, as $1/(k_\text{F}a_\text{s})$ changes continuously from negative to positive, the system exhibits fermion paired superfluity and a BCS-BEC crossover. Extensive studies of BCS-BEC crossover have been made in ultracold atomic gases in the past years.  Recently, many attentions have been paid on studying fermion superfluid and BCS-BEC crossover in the presence of SO coupling. The key emphasis are highlighted below

\vspace{0.1in}

$\circ$ \textbf{Topological Superfluid.} Before the studies in the content of cold atom physics, topological superfluid with SO coupling has been studied extensively in condensed matter systems during recent years. This subject has been reviewed by several articles, such as Ref. \cite{Xiaoliang} and \cite{Alicea}. Kitaev first proposed that a spinless p-wave superconductor in one-dimension can become topological superconductor and exhibit Majorana zero-mode at the edge \cite{Kitaev}. A $p+ip$ superconductor in two-dimension is also a topological superconductor. A conceptual breakthrough was made by a seminal paper Ref. \cite{Fu} which points out $p$-wave interaction is not necessary for a $p$-wave superconductor, and an $s$-wave superconductor can induce topological $p+ip$ pairing in the surface of a topological insulator through proximity effect. Subsequently, Ref. \cite{Sarma, Refael} proposed that a one-dimensional nano-wire with SO coupling and in the presence of Zeeman field can become topological insulator induced by proximity effect of an $s$-wave superconductor. In 2012, the first experiment in nano-wire following this proposal has been reported \cite{Kouwenhoven}. Indeed, signature of zero-bias mid-gap state has been observed \cite{Kouwenhoven}. However, the evidence is not conclusive and whether these mid-gap state is due to Majorana fermions or due to disorder is still under debate.  

Interestingly, the kind of SO coupling with Zeeman field required in this scheme is exactly the Hamiltonian that has been realized with Raman coupling scheme in cold atom setup. One can apply a two-dimensional optical lattice to achieve a strong confinement in the $yz$ plane and realize a one-dimensional gas along $\hat{x}$ direction (i.e. direction of Raman beam and SO coupling). Mean-field studies show that this system indeed exhibits Majorana edge modes at the edge of the cloud or in the domain wall \cite{Jiang,Hu_Majorana,Mueller_Majorana,Zhang_Majorana}. However, the main difference is that in cold atom system, the pairing can naturally originate from interaction between atoms instead of proximity effect. Thus, one concern arises, that is, the pairing order suffers from strong quantum fluctuation in low-dimension and becomes algebraically decayed order, and moreover, there is no true pairing gap in single particle spectrum. It follows the question whether intrinsic attractive interactions can lead to a topological phase and Majorana zero-mode without externally induced pairing. Ref. \cite{Fisher,Sau} discussed this problem and showed that a true long-range order is not essential but two coupled one-dimensional wires with SO coupling is necessary to exhibit protected zero-energy Majorana mode. More recently, Ref. \cite{Altman} reexamined this problem. Because of free from disorder, cold atom system becomes a good candidate to realize such topological superfluid and search for Majorana fermions. However, currently, the major challenge is still from heating problem, as we will discussed at the last session. 

\vspace{0.1in}

$\circ$ \textbf{Richer Pairing Structure.} As discussed in two-body problem, the bound state wave function contains both singlet and triplet component. Similarly, in a many-body problem, the pairing order parameter contains both singlet component and triplet component. Ref. \cite{Melo-1,Melo-3d} solved the BCS pairing problem for Raman-induced SO coupling and found gapless excitations for non-zero $\delta$ in three-dimension. Also, two-body problem has shown that the most favorable momentum for a bound pair may not locate at zero-momentum \cite{Vijay_molecule,Han_molecule}. Consequently, with a mean-field theory for many-body system, Ref. \cite{Wei_FFLO} shows that there is a large region in the parameter space where FFLO state (i.e. pairing with finite momentum) exists stably in two-dimension for non-zero $\delta$. In one-dimension, FFLO also exists for finite detuning $\delta$. Moreover, as discussed above, the superfluid can also exhibit topological properties with SO coupling. Ref. \cite{ChunChen, Liu_Hu_TLFFLO} pointed out that the topological property and FFLO state can coexist which yields a state named as topological FFLO state \cite{ChunChen, Liu_Hu_TLFFLO,Wei_TLFFLO,Chuan_Wei_TLFFLO}. 

\vspace{0.1in}

Finally it is worth to mention that there are also works considering fermion systems with repulsive interaction in the presence of SO coupling. For instance, Ref. \cite{Cui_Ho} studied one-dimensional Fermi gas with infinite repulsion and they find that even small Raman coupling can lead to a dramatic effect in the ground state spin structure.

\section{Rashba Spin-Orbit Coupling}

\subsection{Bosons}

The physical properties share many similarities between bosons with Rashba SO coupling and with Raman-induced SO coupling as discussed above. The major difference is that, with Rashba SO coupling, the degenerate manifold of single-particle ground state forms a continuous circle in the momentum space. This larger degenerate space leads to more intriguing physics in terms of both equilibrium phases and fluctuations. In this section we will emphasize this new aspect. 

\begin{figure}[tb] 
\includegraphics[width=3.5 in]
{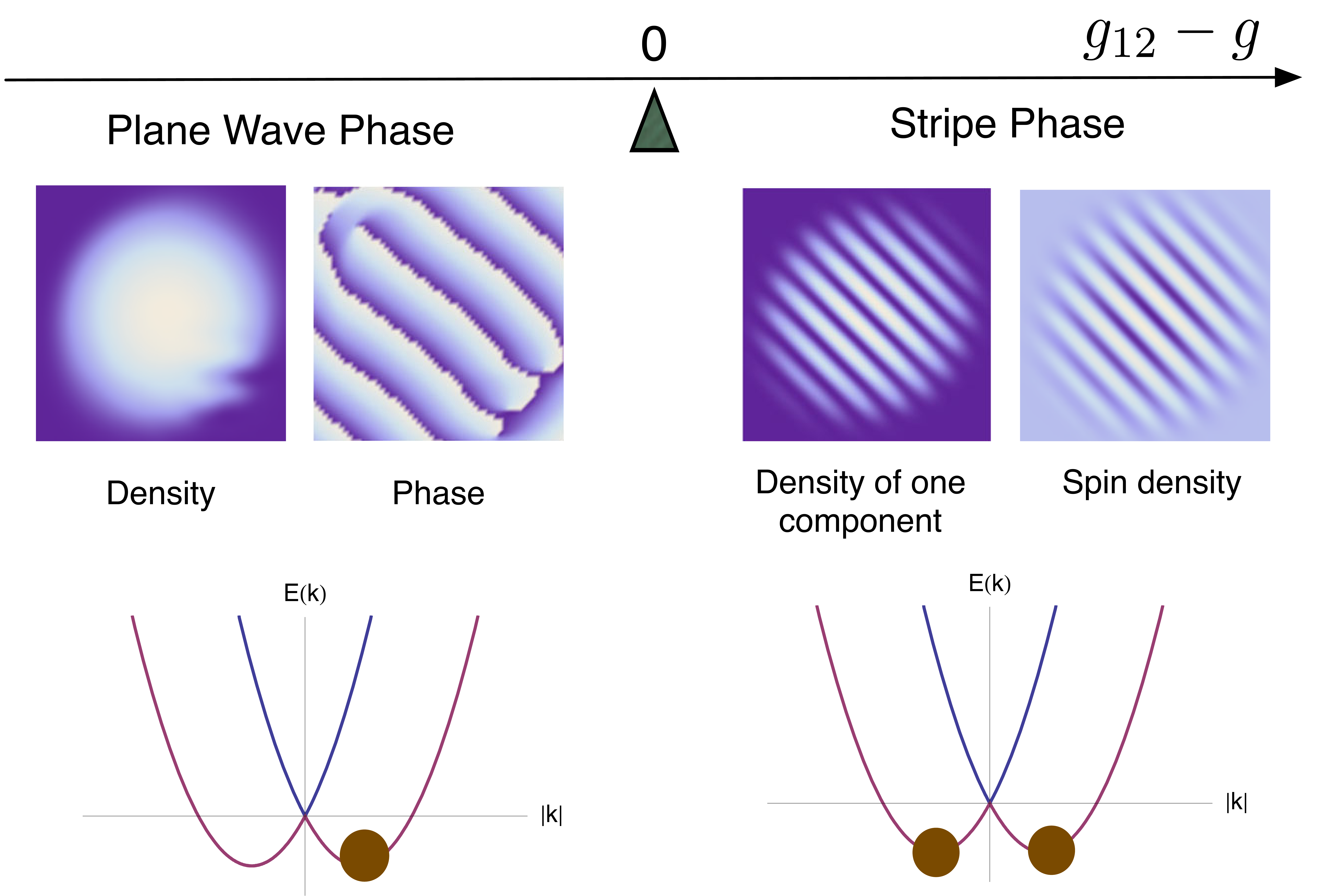}
\caption{Phase diagram as a function of $g_{12}-g$. \label{phase_diagram_Rashba}}
\end{figure} 

\vspace{0.1in}

$\circ$ \textbf{Mean-Field Theory.} The mean-field ground state of Rashba SO coupled bosons is first calculated in Ref. \cite{Zhai_SOCBoson}. The single-particle Hamiltonian is given by Eq. \ref{H_Rashba}, and the single-particle ground states are plane wave states with momentum $\sqrt{k^2_x+k^2_y}=k_0$ and $k_z=0$. A simple form of interaction between two-component bosons is considered as
\begin{equation}
H_\text{int}=\int d^3{\bf r}\left(g n^2_\uparrow +2g_{12} n_\uparrow n_\downarrow +g n^2_\downarrow\right). \label{int_2}
\end{equation} 
Within mean-field theory and by minimizing mean-field energy, it is found that the ground state is determined by the difference between $g_{12}$ and $g$, as summarized in Fig. \ref{phase_diagram_Rashba} \cite{Zhai_SOCBoson}. When $g_{12}<g$, bosons are condensed into a single plane wave state. For instance, if the momentum is along $\hat{x}$ direction, the wave function is given by
\begin{equation}
\Psi=e^{i\varphi}e^{ik_0 x}\frac{1}{\sqrt{2}}\left(\begin{array}{c}1  \\1  \end{array}\right) \label{WF_pw}
\end{equation}
The density is uniform and the phase modulates periodically from zero to $2\pi$ in space.  In additional to spontaneous $U(1)$ phase symmetry breaking, this state also spontaneously breaks the simultaneous rotational symmetry of spin and space in $xy$ plane. When $g_{12}>g$, bosons are condensed into a superposition of two plane wave states with opposite momentums. For instance, if this momentum is along $\hat{x}$ direction, the condensate wave function is given by
\begin{align}
\Psi&=\frac{1}{2}e^{i\varphi}\left(e^{i\theta}e^{ik_0 x}\left(\begin{array}{c}1 \\1\end{array}\right)+e^{-i\theta}e^{-ik_0x}\left(\begin{array}{c}1 \\-1\end{array}\right)\right)\nonumber\\
&=e^{i\varphi}\left(\begin{array}{c}\cos(k_0x+\theta) \\i\sin(k_0 x+\theta)\end{array}\right). \label{stripe_WF}
\end{align} 
Thus, the density of each component modulates in space. Moreover, the high-density regions of one component overlap with the low-density regions of the other component, which represents a microscopic phase separation and therefore the inter-component repulsive interaction energy is minimized. That is why such a state is energetically favorable when $g_{12}$ is larger. This state not only spontaneously breaks the rotational symmetry in the $xy$ plane, but also spontaneously breaks the spatial translational symmetry along one of the spatial direction, which is described by $\theta$ variable. Symmetry wise, it is a superfluid smectic state. Similar results have also been found in spin-1 condensate \cite{Zhai_SOCBoson}. Later it is shown by Ref. \cite{Yu} that when $g_{12}$ is much larger than $g$ and when interaction energy is much larger than SO coupling energy, a phase separation instability toward fully polarized zero-momentum state will also take place.

In the dark-state realization of Rashba SO coupling as discussed in Sec. \ref{Rashba}, the wave function of different ``spin" states in fact have spatial dependence. In this case, interactions between different spin components can not be captured by constant interactions as in Eq. \ref{int_2}. Ref. \cite{Yip,Chuanwei_SOC} considered this complication by using a specific dark-state realization of Rashba SO coupling. The results discussed above remain qualitative unchanged. 

One question is why there is always either one or two momentum components in the ground state wave function. This is because the repulsive interaction in the total density channel favors a uniform total density. Either a plane wave state or a stripe state like Eq. \ref{stripe_WF} can satisfy this requirement. Furthermore, it is easy to show that such a requirement can not be satisfied anymore once there are more than two momentum components.  Then, the next question is whether under certain situation, the condensate wave function will contain more than two momentum components. If this happens, it will give rise to more interesting quantum structures of condensate wave function. The developments along this line can be summarized as follows:

\vspace{0.05in}

1. Spin-2 condensate. Ref. \cite{You_S2,Machida} pointed out that in a spin-2 condensate, in addition to density interaction and spin-interaction (${\bf S}_i\cdot{\bf S}_j$-term), there is an additional term of singlet-pair interaction (i.e. the interaction energy is proportional to the density of single-pair amplitude). This singlet-pair interaction term favors a condensate wave function with four or three different momenta. This gives rise to two-dimensional square or triangular crystals of spin vortices. 

\vspace{0.05in}

2. Condensate inside a tight trap. The presence of harmonic trap breaks the spatial translational symmetry which scatters one momentum to any other momentum state with degenerate energy. If this momentum mixing is strong enough, it would lead to a superposition of all momentum states in the degenerate manifold, which will give rise to topologically distinct state such as half quantum vortex state, as first pointed out by Ref. \cite{Wu, Galitski}. Such a half quantum vortex state breaks time-reversal symmetry and is therefore doubly degenerate. This state is also ground state in absence of interactions. Therefore, as shown in Ref. \cite{Wu,Santos_trap,Hu_trap,Hu_trap_2} by numerically minimizing mean-field energy, when $g_{12}>g$, the system will undergo a transition from stripe phase to a half quantum vortex state, either by increasing harmonic confinement with a fixed interaction strength, or by decreasing interaction with a fixed harmonic confinement.   

\vspace{0.05in}

3. Dipolar condensate. SO coupled condensate with dipolar interactions is first studied in Ref. \cite{SuYi} for Raman-induced SO coupling, where spin vortex is found due to dipolar interactions. For Rashba SO coupling, spin-dependent and spin-independent dipolar interaction are considered by Ref. \cite{Clark} and Ref. \cite{Demler}, respectively. Ref. \cite{Clark} finds that a spin-dependent dipolar interaction plays a similar role as SO coupling in single-particle Hamiltonian, which can lead to a meron spin configuration. Ref. \cite{Demler} finds that a spin-independent interaction favors a superposition of a discrete set of different momenta, and in some parameter region, when the number of momenta are, for instance, five or ten, it will give rise to an interesting quantum quasi-crystal. 

\vspace{0.05in}

Besides, Bose condensate with an three-dimensional isotropic SO as Eq. \ref{H_3d} has also been studied where stable skymion is found in the ground state \cite{Machida_3,Wu_3d}.

\vspace{0.1in}

$\circ$ \textbf{Fluctuations beyond Mean-Field}

Because of the large degeneracy of single-particle ground state, whether mean-field results are reliable becomes questionable, and whether there exists more exotic ground state whose energy is lower than the mean-field state. These are difficult questions, and various approaches have been applied to address these issues from different perspectives. Some of these studies also reveal interesting physics related to excitation spectrum of Rashba SO coupled condensate.  

\vspace{0.05in}

1. Effective Theory Approach: In this approach, we examine the low-energy fluctuation around the mean-field solutions. For stripe phase, there are two low-lying modes which are superfluid phase $\varphi$ and the relative phase $\theta$ corresponding to breaking of spatial translational symmetry. For plane wave phase, superfluid phase $\varphi$ is the only low-lying mode. Following standard method one can integrate out density fluctuations and obtain a low-energy effective theory for low-lying collective modes. For conventional superfluid, such an approach will yield an isotropic $XY$ type theory which gives rise to a linear and isotropic Goldstone mode. Ref. \cite{Jian} applied this approach to SO coupled condensate. For Rashba SO coupled condensate, for stripe phase with the condensate wave function given by Eq. \ref{stripe_WF}, one will obtain an anisotropic $XY$ model for total phase $\varphi$, but for the relative phase $\theta$, the effective Hamiltonian is proportional to $(\partial_x\theta)^2+(\partial^2_y\theta)^2/(4k^2_0)$. This is a consequence of rotational symmetry in $xy$ plane and the energy degeneracy when ordered direction rotates in $xy$ plane. In fact, this Hamiltonian is similar to that of liquid crystals. Such Hamiltonian means the dispersion of low-energy mode is linear along the stripe direction $\hat{x}$, but is soften to be quadratic along the perpendicular direction $\hat{y}$. The topological defect of $\theta$-field is dislocations in the stripe. Such an effective theory for $\theta$ also means that the energy of these dislocations does not logarithmically depend on system size in two-dimension. Therefore, in two-dimension at any finite temperature, thermal fluctuation will create lots of free dislocations in stripe phase and melt the density wave order. The system will enter a paired superfluid phase. For a plane wave phase with wave function given by Eq. \ref{WF_pw}, similar approach shows that the effective Hamiltonian for superfluid phase $\varphi$ behaves as $(\partial_x\varphi)^2+(\partial^2_y\varphi)^2/(4k^2_0)$. In this case, superfluid order will be destroyed at any finite temperature in two-dimension. 

\vspace{0.05in}

2. Renormalized T-matrix approach. The interactions between different momentum states in the degenerate manifold play an essential role in determining the ground state. In momentum space, interaction takes a general form as
\begin{equation}
\hat{H}_\text{int}=\sum\limits_{{\bf k},{\bf k^\prime}\sigma\sigma^\prime}g_{{\bf k}{\bf k^\prime}\sigma\sigma^\prime}\psi^\dag_{{\bf k}\sigma}\psi^\dag_{{\bf k^\prime}\sigma^\prime}\psi_{{\bf k^\prime}\sigma^\prime}\psi_{{\bf k}\sigma}.
\end{equation}
At mean-field level, $g_{{\bf k}{\bf k^\prime}\sigma\sigma^\prime}$ is taken as constants independent of ${\bf k}$ and ${\bf k^\prime}$. Ref. \cite{Sarang, Baym_Tmatrix} considered how the high-order scattering processes renormalize these scattering amplitudes, and argued one should use renormalized interaction to determine the ground state in the dilute limit when chemical potential is much below the energy scale of SO coupling. They find that after renormalization, $g_{{\bf k}{\bf k^\prime}\sigma\sigma^\prime}$ develops strong dependence on angle of ${\bf k}$ and ${\bf k^\prime}$. The repulsive interaction is stronger for ${\bf k}$ and ${\bf k^\prime}$ in the same direction than ${\bf k}$ and ${\bf k^\prime}$ in opposite direction. This means that the renormalization interaction always favors a stripe phase than a plane wave phase. Ref. \cite{Sarang} also reaches the conclusion of a boson pairing instability at finite temperature. This effect can also be understood from a two-body perspective \cite{Qi}.  

\vspace{0.05in}

3. Bogoliubov approach. In absence of interactions, it has been shown that the condensation temperature $T_\text{c}$ is zero with Rashba SO coupling, because of the enhanced low-energy density-of-state \cite{Galitski}. Then next question is whether it is also true in the presence of interactions. It is equivalent to ask whether  the quantum depletion will diverge at any finite temperature. It is found that, surprisingly, condensate fraction remains finite at low-temperature in the presence of interactions \cite{Baym_Tc,Cui_DoS,Liao}. This shows that in contrast to conventional wisdom, interaction stabilizes condensation.  Ref. \cite{Cui_DoS} also finds that for three-dimensional isotropic SO coupling as Eq. \ref{H_3d}, condensate will be all depleted and there will be no Bose condensation at any finite temperature when interaction is spin-independent. 

If the interaction is spin-independent, mean-field energy also can not select out a unique ground state. In this case, one should also consider fluctuation and compare which mean-field ground state has the smallest fluctuation energy. This is order-by-disorder mechanism. Ref. \cite{Wu} and  Ref. \cite{Barnett} studied this question but they obtained opposite results. Ref. \cite{Wu} concludes a stripe state is more favorable while Ref. \cite{Barnett} concludes a plane wave state is more favorable. 

Besides, Ref. \cite{Han_1} develops a hydrodynamic theory for Rashba SO coupled bosons, with which the low-energy excitation modes for both stripe and plane wave states have also been studied. Based on the study of low-energy excitation spectrum, Ref. \cite{Liao_Feynman} discussed whether Feynman relation for excitation dispersion $\omega_{{\bf k}}=\epsilon_0({\bf k})/S({\bf k})$ is still applicable to SO coupled bosons, where $\epsilon_0({\bf k})$ is the free boson dispersion and $S({\bf k})$ is dynamic structure factor. 

\vspace{0.05in}

4. Variational Wave Function. All above approaches are somewhat related to mean-field theory. There are also works that try to find non mean-field ground state. Ref. \cite{Wu,Galitski,Qi} points out various fragmented states as possible ground state, while they also admit that these states are quite fragile and not stable against small perturbations. Ref. \cite{Kamenev} notices that the chemical potential of condensed bosons scales linearly in density $n$, whereas with Rashba SO coupling and in two-dimension, the low-energy density-of-state behaves as $1/\sqrt{\epsilon}$ ($\epsilon$ is energy), similar as a normal one-dimensional system, and therefore, the chemical potential of a noninteracting Fermi gas scales as $n^2$ instead of $n$. Thus in low-density limit a Fermi gas will has lower energy than a Bose gas. Based on this observation, they propose a fermonized many-body state by composite fermion construction as ground state of interacting bosons in dilute limit.

\vspace{0.05in}

Despite of these efforts, the nature of quantum ground state of interacting bosons with Rashba SO coupling remains an open issue. It is also a strong motivation for experimental realization of such a SO coupling in cold atom system, where a strongly correlated quantum state can be expected. 

There are also a series of works related to the effect of harmonic trap. Ref. \cite{Hu_Tc} studies effect of harmonic trap on condensation temperature.  Ref. \cite{Baym_stripe} addresses the issue that in the stripe phase how a spatial oscillating particle flow terminates at the edge of condensate in a harmonic trap. A number of papers have studied single-particle problem with Rashba SO coupling in a harmonic trap, where a Landau-level like spectrum has been found \cite{Santos_trap,Hu_trap,Han_correlated,Wu_topology}, and quantum states with nontrivial topology or with strong correlation have been proposed \cite{Wu_topology,Han_correlated}. SO coupled condensate wave functions have also been studied a rotating harmonic trap. With interaction, it is found that a rotating condensate exhibits quite rich structures by varying trapping potential and interaction parameters, such as skymion lattice and giant vortex \cite{Han, Wu_2,Liu,Mason}.

\subsection{Fermions}

Most theoretical studies of Rashba SO coupled Fermi gas focus on strongly interacting regime nearby a Feshbach resonance. The BEC-BCS crossover with Rashba SO coupling is first studied by Ref. \cite{Shenoy_MF,Chuanwei_MF,Pu_MF,Yu_MF}. For this discussion, the main difference between Rashba SO coupling and Raman-induced SO coupling is that the low-energy DoS is significantly enhanced for Rashba case, because of which two-body bound state exists for any weak attractive interactions \cite{Vijay_2body}. In fact, in some earlier study in the content of electron gases, it has already been pointed out that superconductivity can be enhanced because of larger DoS at Fermi surface, when the strength of SO coupling exceeds Fermi energy \cite{Marsiglio}. Recent studies reveal that the bound state physics plays a more important role, in particular, when the strength of SO coupling much exceeds Fermi energy, and the size of bound state becomes much smaller than inter-particle distance. In this regime, the many-body physics is largely determined by Bose condensation of these bound pairs. Previously in three-dimension without SO coupling, this only happens at the BEC side with positive scattering length $a_\text{s}$. While with SO coupling, this can also happens in the BCS side with negative scattering length, where bound state does not exist in absence of SO coupling. In the BCS side, the size of bound pairs decreases with the increasing of SO coupling strength. Thus, even with negative scattering length, the physics can be driven to BEC-type by increasing SO coupling \cite{Shenoy_MF,Chuanwei_MF,Pu_MF,Yu_MF}. Effects of such a bound state have also been considered in attractive interacting Bose gas \cite{Yin}.

\begin{figure}[tb] 
\includegraphics[width=3.5 in]
{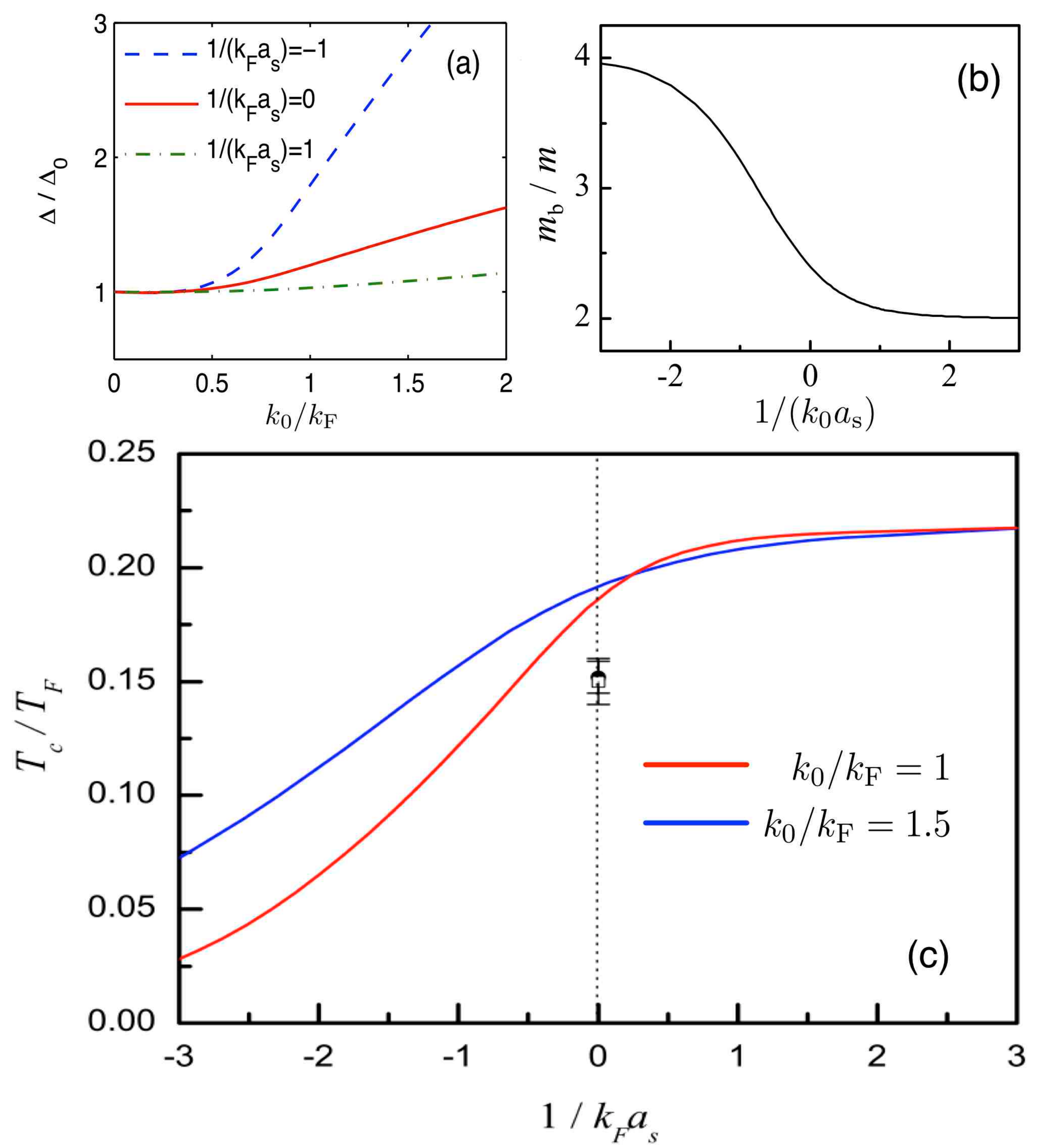}
\caption{(a) Pairing gap as a function of SO coupling strength $k_0/k_\text{F}$ for different interaction strengths. (b) Mass of molecule as a function of $1/(k_0a_\text{s})$. (c) Estimation of superfluid transition temperature as a function of $1/(k_\text{F}a_\text{s})$ for $k_0/k_\text{F}=1$ and $k_0/k_\text{F}=1.5$, respectively. Reprinted from Ref. \cite{Yu_MF}  \label{pairing_SOC}}
\end{figure} 

It is therefore clear that the effect of Rashba SO coupling is stronger in the BCS side than in the BEC side, as shown in Fig. \ref{pairing_SOC}(a). At the BCS side, as well as in the unitary regime, pairing gap increases rapidly when $k_0$ is of the order of $k_\text{F}$, whereas in the BEC side, pairing gap is not sensitive to $k_0/k_\text{F}$. In addition, the mass of molecules is anisotropic and its dispersion behaves as $E_\text{b}({\bf q})=E_0+q^2_\perp/(2m_\text{b})+q^2_z/(4m)$. $m_\text{b}/m$ is a function of $1/(k_0a_\text{s})$ as shown in Fig. \ref{pairing_SOC}(b). In the positive $a_\text{s}$ side $m_\text{b}$ equals $2m$ while it approaches $4m$ as $k_0a_\text{s}\rightarrow 0^-$.  For negative $a_\text{s}$ and sufficiently large $k_0/k_\text{F}$, the superfluid transition temperature is also mostly determined by condensate temperature of bound pairs. Therefore, $T_\text{c}$ can reach the same order as $T_\text{F}$. As shown in Fig. \ref{pairing_SOC} (c), from an estimation based on thermal fluctuation of pairs, one can see that $T_\text{c}$ can arise to about $0.1T_\text{F}$, say, $1/(k_\text{F}a_\text{s})\approx -2$, when interaction is not sufficiently strong \cite{Yu_MF}. This high transition temperature has later been confirmed by more serious $T$-matrix calculation \cite{He_Tc}. If in term of temperature scale of electron gases, this is a very high superconductivity transition temperature. Therefore, this study reveals an alternative route toward searching for high temperature superconductor. 

Along this line, many works have carried out detailed studies of condensate fraction \cite{Salasnich,Zhou_fermion,He_1}, superfluid fraction \cite{Zhou_fermion,He_1}, collective modes \cite{Vijay_modes,He_3},  interaction between bound pairs \cite{Vijay_modes}, Kosterlitz-Thouless transition temperature in two dimension \cite{He_1,Melo_2}, and a number of thermal dynamic properties with mean-field \cite{Melo_1} and pairing fluctuation theory \cite{Han_1}. Similar studies have also been extended to three-dimensional isotropic SO coupling of Eq. \ref{H_3d} \cite{He_2}.  

A large number of papers have treated fermion superfluid with both Rashba SO coupling and Zeeman field, or equivalently, population imbalance \cite{TSF_Zhang,Sato,Iskin_1,Wei,Yi_2,Zhang_T,Yi_3,Liu_impurity,Dong_FFLO,Hu_FFLO,Yi_FFLO,Zhang_FFLO, Zhang_FFLO_lattice,Wei_TLFFLO,Chuan_Wei_TLFFLO}. The issues here are two fold. First, without SO coupling, it has been extensively discussed before that increasing Zeeman field or population imbalance will drive a transition from superfluid to normal state, and possibly a FFLO state also exists in small parameter space. The question is how SO coupling affects this transition and the existence of FFLO phase. Secondly, such a model supports topological superconductor phase \cite{Fu}. Similar topological superfluid phase also exists in cold atom setup \cite{TSF_Zhang,Sato}. The question is how to optimize the parameter space for topological superfluid phase and how to detect this phase in cold atom systems.  

Mean-field phase diagrams with population imbalance have been calculated by Ref. \cite{Iskin_1,Wei,Yi_2}. Some address the stability of fermion pairing against population imbalance in the presence of SO coupling, and some address the regime for topological phase in a harmonic trap. Beyond mean-field level, finite temperature effect on topological phase has also been studied in Ref. \cite{Zhang_T} by including Gaussian fluctuations, and polaron to molecule transition has been studied by Ref. \cite{Yi_3} by variational wave function approach. Ref. \cite{Liu_impurity} studied impurity in topological superfluid and suggests it as a detection method for topological superfluid. As for FFLO state, it has been revealed that FFLO state is much more favorable in the presence of SO coupling \cite{Dong_FFLO,Hu_FFLO,Yi_FFLO,Zhang_FFLO,Zhang_FFLO_lattice}. Moreover, similar as in Raman-induced SO coupling case, it has been shown that the finite momentum FFLO pairing can also coexist with nontrivial topological properties \cite{Wei_TLFFLO,Chuan_Wei_TLFFLO}.

\section{Challenges and Future Directions}

In the near future, there will be further developments in the directions mentioned above, since many issues still require more theoretical and experimental efforts. For instance, the ground state of bosons with Rashba SO coupling may display strong correlations and this still remains as an unsolved problem so far. In addition, there are new directions that have not been well studied so far, but definitely deserve more attentions. I will list a few of them below:

\vspace{0.1in}

$\circ$ \textbf{Spin-Orbit Coupling with Optical Lattices.} It will be interesting to study the interplay between SO coupling and various strongly correlated phases in lattice systems. Boson Hubbard model with Rashba SO coupling has been studied by several works \cite{Trivedi,Galitski_lattice,Wu_lattice,Sengupta,Lewenstein_lattice,Peotta}. They studied how SO coupling affects the superfluid-Mott insulator transition \cite{Trivedi,Sengupta,Lewenstein_lattice} and discussed rich magnetic orders in the Mott insulator phase due to Dzyaloshinskii-Moriya interaction from SO coupling \cite{Trivedi,Galitski_lattice, Wu_lattice,Peotta}. Transport phenomena also show nontrivial physics in both Mott insulator \cite{Duine_lattice} and condensate phase \cite{Cooper_lattice}, because SO coupling can give rise to nontrivial Berry curvature in momentum space for energy bands \cite{Duine_lattice,Cooper_lattice}. With fermions, it is more close to simulating physics related to topological insulator \cite{Chip,Hofstetter,Hur}. In Raman-induced SO coupling scheme, it has been shown experimentally for both bosons and fermions that by applying an additional radio-frequency field, a lattice potential will be generated on top of SO coupling \cite{MIT_SOC,Spielman_lattice}. More interestingly, by tuning the ratio between Raman-coupling and radio-frequency strengths, a band with quite flat dispersion will emerge in a large parameter regime \cite{Zhang_lattice,Scarola}. Here the emergent of flat band is a cooperative effect of SO coupling and lattice effect. In this regime, interactions play a very important role and will lead to many interesting physics. So far, some works have studied condensate stability \cite{Zhang_lattice,Linder}, strongly correlated phases of bosons and fermions \cite{Scarola} in this regime.

Optical lattice also provides an alternative route for implementing SO coupling effect. Recently several experiments have succeeded in inserting large flux per unit cell into a square optical lattice  \cite{Bloch_stagger_flux,Bloch_flux,Ketterle_flux}. It is promising that one can generalize this scheme to realize a situation that different spin component experiences an opposite magnetic field, and such a spin-dependent magnetic field also give rise to SO coupling and quantum spin Hall effect \cite{Bloch_flux,Ketterle_SOC}. Another development relates to engineering band structure in optical lattices. With shaking optical lattice technique, the band dispersion can be changed from single-minimum to double-minimum, as demonstrated in recent experiments \cite{Sengstock1,Cheng}. This is closely related to effects from SO-coupling-modified dispersions discussed above, for instance, superfluid with $Z_2$ symmetry breaking is one of such example \cite{Zheng_Z2}.

\vspace{0.1in}

$\circ$ \textbf{Few-Body Problem with Spin-Orbit Coupling.} Few-body problem is of great interest and importance in cold atomic gases. First, few-body physics itself displays many interesting features such as Efimov states with universal scaling. Secondly, since cold atom is a dilute system, many basic properties of this quantum gases, such as its lifetime due to atomic loss, is determined by few-body physics process. Thirdly, solutions from few-body problem also provides insight to understand correlations in many-body system. Nevertheless, except for extensive studies of two-body problems with SO coupling, as discussed in Sec. \ref{TB}, only two of recent works have studied three-body problems in the presence of SO coupling \cite{Zheyu,Cui_few}. Ref. \cite{Zheyu} discovered that a three-dimension isotropic SO coupling will enhance the formation of universal trimers, and Ref. \cite{Cui_few} pointed out universal borromean three-body binding exists for Rashba SO coupled Fermi gas due to the circle of degenerate in the lowest energy. These studies indicate that much richer physics in this new few-body system is remained to be discovered, and these studies will help to understand the properties of this gas and its many-body physics better. 

\vspace{0.1in}

$\circ$ \textbf{Spin-Orbit Coupled High Spin Systems.} In condensed matter system, research on SO coupling effects is limited to spin-$1/2$ particle of electrons. In recent studies of SO coupling in atomic gases, most theoretical work and almost all experimental works also focused on spin-$1/2$ system. However, one unique feature of atomic quantum gas is that most atoms have high spins.  The conventional wisdom from condensed matter studies is that the larger spin system is more classical because the fluctuation is weaker. However, it is not the case in cold atom system \cite{Wu_large_spin}. For a spin-$S$ atom, a full quantum mechanics description of its wave function requires $2S$ points in the Bloch sphere known as Majorana representation \cite{non-abelian,Schwinger,highspins,highspins2}, and for large spin atom like Dy ($S=8$), the wave function can have nonabelian symmetry represented by various point groups of sixteen points distributed in the unit sphere \cite{Biao}. Viewing large internal degree of freedom as another synthetic dimension, SO coupling can also introduce large synthetic magnetic field in the synthetic space \cite{dimension}. SO coupling in this high spin bosons or fermions systems is expected to exhibit novel quantum phenomena \cite{Zhai_Dy}. Very recently, the first experiment of SO coupled spin-1 gas has been reported from NIST group, and SO coupled Dy experiment is going on in Stanford group. SO coupling experiment on alkali-earth-(like) atom like Yb and Sr with large nuclear spins have also been considered by serval groups. These new experimental efforts will inspire more progresses along this direction.

Finally I would like to mention that the major challenge now is the heating problem. So far all experiments are performed with alkali atoms. For alkali atoms, the spin flipped Raman process is proportional to $W\Delta_\text{FS}/\Delta^2$, where $W$ is laser power, $\Delta_\text{FS}$ is the fine structure splitting of excited states, and  $\Delta$ is the detuning, as discussed in Sec. \ref{Raman_SO}. On the other hand, heating from spontaneous emission also depends on $W\Gamma/\Delta^2$, where $\Gamma$ is the line width of excited states. Since the heating rate and Raman coupling scale in the same way with respect to the ratio of $W$ to $\Delta$, it is hard to suppress heating with fixed Raman coupling. This is in contrast to laser trapping and cooling case, where only scalar atom-light coupling is used. The scalar process scales as $W/\Delta$, and therefore one can suppress heating by increasing detuning to far detuned regime. In the Raman coupling case, the smaller fine structure coupling $\Delta_\text{FS}$, the stronger heating effect for same Raman coupling strength.  For instance, the heating effect in ${}^6$Li experiment \cite{MIT_SOC} is much stronger than in ${}^{40}$K experiment \cite{Jing_SOC}. Even in ${}^{40}$K experiments, it has been observed that the temperature is increased from $0.2T_\text{F}$ to above $0.5T_\text{F}$ or even higher, when Raman coupling is turned on for certain time \cite{Jing_SOC}. This precludes observation of many interesting many-body physics, in particular, in fermion systems. 

One solution is to utilize other class of atoms. One candidate is alkali-earth or alkali-earth-like atom (such as Yb), in which the electronic ground state and one metastable electronic excited state form a spin-$1/2$ manifold and they are directly coupled by single laser light. However, in this case the collisional stability of excited states is a concern. Other candidate is more complicated open shall lanthanide atoms such as Dy and Er. In these atoms, atomic spin in their ground state has nonzero angular momentum component ${\bf L}$ and laser can directly flip spin, so that the Raman coupling strength depends on $W/\Delta$ as usual scalar process. Some excited states of these atoms also have very small line width. Putting all these effects together, the heating rate can be reduced by several orders of magnitudes for same Raman coupling strength comparing to alkali atoms \cite{Zhai_Dy}. Besides, there are also proposals to use pure magnetic schemes to generate SO coupling, as discussed in Sec. \ref{Rashba}, where heating from spontaneous emission can be avoided. All these proposals offer promises for future studies. 

{\it Acknowledgements.} I would like to thank Zeng-Qiang Yu, Xiaoling Cui, Wei Zheng, Chao-Ming Jian, Zhu Chen, Zheyu Shi, Tin-Lun Ho, Benjamin Lev, Biao Lian, Shizhong Zhang, Peng Zhang, Chao Gao, and in particular, Shanxi University experimental group led by Jing Zhang and USTC experimental group led by Shuai Chen and Jian-Wei Pan, for very fruitful collaborations on this topic during the past a few years. This work is supported by Tsinghua University Initiative Scientific Research Program,  NSFC under Grant No.  11174176 and 11325418, and NKBRSFC under Grant No. 2011CB921500.

\end{document}